\documentclass[aps,prb,reprint,superscriptaddress,amsmath,amssymb,showpacs]{revtex4-2}%

\newcommand{\A}{\mathscr{A}}

\newcommand{\C}{\mathscr{C}}
\newcommand{\LL}{\mathcal{L}}
\newcommand{\LS}{\mathscr{L}}
\newcommand{\RR}{\mathcal{R}}

\newcommand{\QQ}{\mathcal{Q}}

\newcommand{\dd}{\mathrm{d}}

\newcommand{\ii}{\text{i}}

\newcommand{\tr}[2]{\text{Tr}_{ #1 } \left\{ #2 \right\}}

\newcommand{\av}[1]{\langle #1 \rangle}
\newcommand{\avg}[2]{\langle #1 \rangle_{#2}}
\newcommand{\expo}[1]{\text{exp}\left( #1 \right)}
\newcommand{\stat}{\text{st}}
\newcommand{\tS}{\text{S}}
\newcommand{\tL}{\text{L}}
\newcommand{\tR}{\text{R}}
\newcommand{\tI}{\text{I}}

\newcommand{\tB}{\text{B}}
\newcommand{\tSB}{\text{SB}}

\newcommand{\tot}{\text{tot}}

\newcommand{\e}{\text{e}}

\newcommand{\unit}[1]{\,\mathrm{#1}}

\newcommand{\dket}[1]{\ket{ #1 } \rangle }
\newcommand{\dbra}[1]{\langle \bra{ #1 } }
\newcommand{\dbraket}[1]{\langle \braket{ #1 } \rangle }

\newcommand{\pd}[2]{\frac{\partial #1}{\partial #2}}  
  
\usepackage{braket}
\usepackage{subfigure}
\usepackage{graphicx}
\usepackage{xcolor}
\usepackage{dsfont}
\usepackage{pstricks}
\usepackage{mathrsfs}
\usepackage{bbm}

\graphicspath{{./overview_5e-2g_06lo/}{../../../HEOM/programs/el_vib_bath/adiabatic_regime/Plots/}{../../../HEOM/programs/system_vib_pola_trafo/Rainer_PHD_model/Plots/}{../../../HEOM/programs/Pade_decomposition/Results/}{./Plots_mm_F_V/}{./Plots_1m_fig/}{./Plots_BDBT/}}

\begin{document}

\title{Hierarchical quantum master equation approach to current fluctuations in nonequilibrium charge
transport through nanosystems}

\author{C.\ Schinabeck}
\affiliation{Institute for Theoretical Physics and Interdisciplinary Center for Molecular Materials, \\
Friedrich-Alexander-Universit\"at Erlangen-N\"urnberg,\\
Staudtstr.\, 7/B2, D-91058 Erlangen, Germany
}
\affiliation{Institute of Physics, University of Freiburg, Hermann-Herder-Strasse 3, D-79104 Freiburg, Germany}
\author{M.\ Thoss}
\affiliation{Institute of Physics, University of Freiburg, Hermann-Herder-Strasse 3, D-79104 Freiburg, Germany}

\date{\today}

\begin{abstract}
We present a hierarchical quantum master equation (HQME) approach, which allows the numerically exact evaluation of higher-order current cumulants in the framework of full counting statistics for nonequilibrium charge transport in nanosystems.
The novel methodology is exemplarily applied to a model of vibrationally coupled electron transport in a molecular nanojunction. We investigate the influence of cotunneling on avalanche-like transport, in particular in the nonresonant transport regime, where we find that inelastic cotunneling  acts as trigger process for resonant avalanches.
In this regime, we also demonstrate that the correction to the elastic noise upon opening of the inelastic transport channel is strongly affected by the nonequilibrium excitation of the vibration as well as the polaron shift.
\end{abstract}


\maketitle
In order to characterize and understand electron transport in nanosystems, not only the average current but also its fluctuations and higher-order cumulants of the distribution of transferred charge, the so-called full counting statistics (FCS) \cite{Levitov1993,Lee1995,Levitov1996}, are of great interest and can be determined in experiments \cite{Reulet2003,Bomze2005,Gustavsson2006,Ubbelohde2012}.
The higher-order current cumulants provide important information, e.g.\ the study of the current noise (second cumulant) has allowed the determination of the charge of quasiparticles in a superconductor \cite{de_Jong1994,Jehl2000,Lefloch2003} and quantum Hall systems \cite{Picciotto1997,Saminadayar1997} as well as of the number and transparency of transmission channels in nanosystems \cite{Brom1999,Cron2001,Djukic2006,Kiguchi2008,Tal2008,Wheeler2010,Schneider2010,Chen2012}.

While the FCS in noninteracting systems is described by the Levitov-Lesovik formula \cite{Levitov1996,Nazarov2009}, the theoretical evaluation of higher-order cumulants is challenging in interacting systems.
To this end, different approximate methods have been applied including master equation \cite{Hershfield1993,Korotkov1994,Koch2006,Haupt2006,Flindt2008,Emary2009,Esposito2010,Schinabeck2014,Kaasbjerg2015,Agarwalla2015,Kosov2017,Rudge2019} and nonequilibrium Green's function approaches \cite{GalperinShot2006,Haug2008,Avriller2009,Schmidt2009,Haupt2010,Novotny2011,Park2011,Utsumi2013,Agarwalla2015,Souto2015,Miwa2017,Dong2017,Tang2017,Stadler2018}, which are either perturbative in nature or rely on special factorization schemes.
Although interesting transport phenomena have been revealed by these approximate methods, a reliable and quantitative analysis, in particular in systems with strong coupling, requires the application of methods, which can be systematically converged, i.e.\ numerically exact methods.
In this context, for example, Cohen and coworkers have computed the FCS of the nonequilibrium Anderson impurity model with a quantum Monte-Carlo method \cite{Ridley2018}. 

In this paper, the hierarchical quantum master equation (HQME) approach is extended to the evaluation of the FCS for nonequilibrium charge transport.
The HQME approach generalizes perturbative master equation methods by including high-order contributions as well as non-Markovian memory and allows for the systematic convergence of the results \cite{Haertle2013a,Haertle2015}.
The method (also called hierarchical equation of motion (HEOM) approach) was originally developed by Tanimura and Kubo to study relaxation dynamics \cite{Tanimura1989,Tanimura2006}. Yan and coworkers have extended it to fermionic charge transport \cite{Jin2007,Jin2008}. In this context, they \cite{Zheng2009,Li2012,Zheng2013,Cheng2015,Ye2016,Cheng2017,Li2017,Hou2017,Hou2017a} as well as H\"artle \emph{et al.} \cite{Haertle2013a,Haertle2014,Haertle2015,Wenderoth2016} have investigated electron-electron interactions. We have recently proposed two formulations to treat electronic-vibrational interaction with the HQME approach \cite{Schinabeck2016,Schinabeck2018}.
To extend the methodology to the evaluation of current fluctuations within the framework of FCS, a counting field is added to the HQMEs and the approach of Flindt \emph{et al.} \cite{Flindt2008,Flindt2010} is employed in the hierarchical Liouville space \cite{Li2012,Wang2013} to obtain the higher-order current cumulants beyond the average current.

We exemplarily apply the novel theoretical methodology to study current fluctuations in nanosystems with strong coupling to mechanical or vibrational degrees of freedom. This is the case, for example, in single-molecule junctions \cite{Tal2008,Secker2011,Neel2011,Lau2016} and suspended carbon nanotubes \cite{Weig2004,Sapmaz2006,Leturcq2009}, where  strong electronic-vibrational interaction leads to interesting nonequilibrium transport effects, such as switching \cite{Molen2010}, negative differential resistance \cite{Gaudioso2000,Pop2005,Galperin2005,Leijnse2008,Haertle2011}, enhanced current fluctuations \cite{Koch2005,Koch2006,Secker2011,Schinabeck2014}, decoherence \cite{Haertle2011b,Ballmann2012}, and local heating or cooling \cite{Pop2005,Koch2006a,Haertle2011,Haertle2011c,Haertle2015a,Gelbwaser2018,Haertle2018}.
Employing the HQME approach, we investigate the influence of cotunneling on avalanche-like transport, in particular in the nonresonant transport regime. In this regime, we also demonstrate that the sign of the inelastic correction to noise at the opening of the inelastic transport channel is strongly influenced by the nonequilibrium vibrational excitation.  
To be specific, we adopt in the following the terminology used in the context of charge transport in molecular junctions. It should be noted, though, that the
methodology is as well applicable to other nanosystems with
strong electronic-vibrational coupling as mentioned above.

We consider a model for vibrationally coupled electron transport described by the Hamiltonian $H=H_\tS + H_\tSB + H_\tB$
with ($\hbar=1$)
 \begin{align}
 \begin{split}
  H_\tS = & \sum_m \epsilon_m d^\dagger_m d_m + \sum_\alpha \Omega_\alpha a^\dagger_\alpha a_\alpha\\
  &+ \sum_{m,\alpha} \lambda_{m\alpha} d^\dagger_m d_m (a^\dagger_\alpha + a_\alpha),
  \end{split}
 \end{align}
$H_{\tSB} = \sum_{k \in \tL/\tR, m} ( V_{k m} d^\dagger_m c_k + h.c. )$, and $ H_\tB = \sum_{k \in \tL/\tR} \epsilon_k c^\dagger_k c_k$.
Having performed a system-bath partitioning, the molecular subspace ($H_\tS$) is considered as reduced system while the leads form the bath ($H_\tB$).
The molecule is modeled by a set of electronic levels with energy $\epsilon_m$, which are coupled to vibrational modes with frequencies $\Omega_\alpha$ via coupling constants $\lambda_{m \alpha}$ and interact with a continuum of electronic states in the leads ($K=\tL,\tR$), described by the spectral density 
$ \Gamma_{K,mm'}(\omega) = 2 \pi \sum_{k \in K} V_{km}^* V_{km'} \delta( \omega -\epsilon_k)$.

To obtain the FCS $P_\tR(t,q)$ in the steady state regime, the number $q$ of electrons is counted, which has been collected in the right lead in the time span $[0,t]$.
Based on a two-point projective measurement scheme \cite{Esposito2009}, the characteristic function is given by the Fourier transform with respect to $q$,
\begin{align}
G_\tR(t,\chi)= \sum_{q} P_\tR (t,q) \e^{\ii \chi q}= \tr{\tS + \tB}{\rho_\tot(t,\chi)},
\label{eq:FCS_charact_func_rho}
\end{align}
where $\chi$ is referred to as the counting field and $\rho_\tot(t,\chi)$ denotes the density operator of the overall system dressed by $\chi$.
In the long time limit, the time derivative of the cumulants
$C^r_\tR(t) = \partial{^r}/\partial(\ii \chi)^r \ln G_\tR(t,\chi)$
corresponds to the steady state zero-frequency current cumulants,
$\braket{\braket{I_\tR^r}} = \lim_{t \to \infty} \pd{}{t}\  C^r_\tR(t)$.
The first two current cumulants are the average current $\av{I} \equiv \braket{\braket{I_\tR^1}}$ and the zero-frequency current noise $S(\omega=0) \equiv \braket{\braket{I_\tR^2}}$.
The dressed density operator $\rho_\tot(t,\chi)$ can be written as \cite{Esposito2009}
\begin{align}
 \rho_\tot(t,\chi)= U_\chi(t,0) \rho_\tot(0) U_{-\chi}^\dagger (t,0),
\label{eq:red_dens_U_chi} 
\end{align}
where $U_\chi(t,0)=\check{T} \expo{-\ii \int_0^t \dd \tau (H_\tS + H^\tI_{\tSB,\chi} (\tau))}$ denotes the dressed time-ordered propagator in the bath interaction picture with $H^\tI_{\tSB,\chi}(t)=\e^{\ii H_\tB t} H_{\tSB,\chi} \e^{-\ii H_\tB t} $.
To add the counting field $\chi$ to the Hamiltonian, we apply the transformation $H^\tI_{\tSB,\chi}(t) = \e^{\ii (\chi / 2) N_\tR} H_\tSB^\tI(t) \e^{-\ii (\chi / 2) N_\tR}$ \cite{Esposito2009}.

Following the standard derivation of the HQME method \cite{Jin2008}, the following set of EOMs dressed by the counting field $\chi$ is obtained,
\begin{align}
\begin{split}
&\pd{}{t} \rho^{(n)}_{j_n \cdots j_1 }(t;\chi) =- \left( \ii \LL_\tS + \sum_{r=1}^{n}  \gamma_{j_r} \right) \rho^{(n)}_{j_n \cdots j_1 }(t;\chi)\\
&- \ii \sum_{r=1}^{n} (-1)^{n-r}\ \C_{j_r} \rho^{(n-1)}_{j_{n} \cdots j_{r+1} j_{r-1} \cdots j_1} (t;\chi)\\
&- \ii \sum_{K,\sigma,l} \left( \A^{\bar \sigma}_{0,K} + \A^{\bar \sigma}_{1,K} \e^{- \sigma \ii \chi \delta_{K,\tR} } \right) \rho^{(n+1)}_{(0,K,\sigma,l) j_n \cdots j_1} (t;\chi)\\
&+ \ii \sum_{K,\sigma,l} \left( \A^{\bar \sigma}_{0,K} \e^{+ \sigma \ii \chi \delta_{K,\tR} } + \A^{\bar \sigma}_{1,K} \right) \rho^{(n+1)}_{(1,K,\sigma,l) j_n \cdots j_1} (t;\chi),
\end{split}
\label{eq:HQME_chi}
\end{align}
with
\begin{subequations}
\label{eq:HQME_chi_CA}
\begin{align}
  \C_{0,K,\sigma,l}\; \rho^{(n)}(t;\chi) =& \eta_{l} \sum_m V_{K,m} d_m^\sigma\; \rho^{(n)}(t;\chi), \\
  \C_{1,K,\sigma,l}\; \rho^{(n)}(t;\chi) =& (-)^{n} \eta^{*} _{l}\; \; \rho^{(n)}(t;\chi) \sum_m V_{K,m} d_m^\sigma, \\
 \A^{\bar \sigma}_{0,K}\; \rho^{(n)}(t;\chi)=& \sum_m V_{K,m} d_m^{\bar \sigma} \; \rho^{(n)}(t;\chi),\\
 \A^{\bar \sigma}_{1,K}\; \rho^{(n)}(t;\chi)=& (-)^{n} \rho^{(n)}(t;\chi)\; \sum_m V_{K,m} d_m^{\bar \sigma}.
\end{align}
\end{subequations}
As the forward and the backward propagator in Eq.\ \eqref{eq:red_dens_U_chi} carry a different counting-field dependence, the respective contributions are separated in the HQMEs \eqref{eq:HQME_chi} by the index $f=0,1$ in the multiindex $j=(f,K,\sigma,l)$.
Consequently, the ADOs $\rho^{(n)}_{j_n \cdots j_1 }(t;\chi)$ fulfill the relation $\sum_{f_n \cdots f_1=0,1} \rho^{(n)}_{j_n,\ldots,j_1} = \rho^{(n)}_{p_n,\ldots,p_1}$,
where $\rho^{(n)}_{p_n,\ldots,p_1}$ with $p = (K,\sigma,l)$ denote the ADOs of the standard HQME formulation.

The index $l$ originates from the decomposition of the correlation function of the free bath $C^\sigma_{K,mm'}(t) = \avg{ \breve c_{Km}^\sigma(t) \breve c_{Km'}^{\bar \sigma} (\tau) }{\tB}$ by a sum over exponentials, i.e.\ $C^\sigma_{K,mm'}(t) = V_{Km} V_{Km'} \sum_{l=0}^{l_\text{max}} \eta_{l} \e^{-\gamma_{K,\sigma,l} t}$, with the bath coupling operators $\breve c^\sigma_{Km} (t) = \e^{\ii H_\tB t} \left( \sum_{k \in K} V^{\bar \sigma}_{km} c^\sigma_{k} \right) \e^{-\ii H_\tB t}$.
To this end, the spectral density in the leads is modeled as a single Lorentzian
$\Gamma_{K,mm'} (\omega)= 2 \pi \frac{V_{Km} V_{Km'} W^2}{(\omega-\mu_K)^2 +W^2}$
and the Fermi distribution is approximated by a sum-over-poles scheme, the so-called Pade decomposition. To mimic the wide-band limit, the band width is chosen as $W=10^4 \unit{eV}$.
\par
As the HQMEs \eqref{eq:HQME_chi} are local in time, the recursive scheme proposed by Flindt \emph{et al.} \cite{Flindt2008,Flindt2010}, which was originally developed for a lowest-order Markovian ME in the Liouville space of the reduced system, can be employed to evaluate the zero-frequency current cumulants.
To this end, the EOMs \eqref{eq:HQME_chi} are represented in matrix-vector form in the hierarchical Liouville space \cite{Li2012,Wang2013}, $\ket{\dot{\rho} (t;\chi)} \rangle_H = \LS_H (\chi) \ket{\rho (t;\chi)} \rangle_H$
where the column vector $\ket{ \rho (t;\chi) } \rangle_H$ contains all ADOs included in the calculation.
Following the recursive scheme of Flindt \emph{et al.}, the first two zero-frequency current cumulants are expressed by $\dbraket{I_\tR^1} = e\dbraket{\tilde 0| \LS^{(1)}_H |0}$ and
\begin{align}
 \dbraket{I_\tR^2} =& e^2\dbraket{\tilde 0| \LS^{(2)}_H |0} - 2 e^2 \dbraket{\tilde 0| \LS^{(1)}_H \RR \LS^{(1)}_H |0},
\end{align}
where $\dket{0} \equiv \dket{\rho^\stat}$ and $\dbra{0} \equiv \dbra{ \mathbbm{1}_\tS,0,\ldots,0}$ denote the left and the right eigenvector of $\LS_H (\chi=0)$ with eigenvalue 0, respectively.
Additionally, we have defined $\LS^{(r)}_H \equiv \partial^r / \partial (\ii \chi)^r \LS_H (\chi) |_{\chi=0}$ and the pseudoinverse $\RR=\QQ\LS^{-1}_H \QQ$ with $\QQ =1-\dket{0} \dbra{\tilde 0}$.
In case of a noninteracting system $(\lambda_{m\alpha}=0)$, converged results for the current (noise) representing a single-particle (two-particle) observable are obtained by a truncation of the hierarchy at the second (fourth) tier.
Cerrillo \emph{et al.} have proposed an alternative HQME method to evaluate the FCS for the case of energy transfer in the spin-boson model where a cascade of hierarchies is set up for the moments of the FCS \cite{Cerrillo2016}. This approach could, in principle, be reformulated for fermionic charge transport. However, it cannot be directly solved for the steady state in contrast to our approach outlined above.
The frequency-dependent current noise can be calculated following Ref.\ \onlinecite{Jin2015}.

%
In the following, we apply the methodology to study current fluctuations in nanojuntions with strong electronic-vibrational coupling. A particular intriguing nonequilibrium phenomenon in this parameter regime is the existence of avalanche-like transport with giant Fano factors.
This was studied previously employing 
perturbative Markovian rate equations \cite{Koch2006,Schinabeck2014}.
Here we use the nonperturbative HQME approach to investigate the influence of cotunneling on avalanche-like transport.
In order to quantify the noise, the Fano factor is used, which is defined as the ratio between the actual zero-frequency noise and the noise of a Poissonian process, i.e.\ $F=S(0) / e \av{I}$.

\begin{figure}[b!]
	\centering
	\begin{tabular}{ll}
	  \includegraphics[width=.49\columnwidth]	 {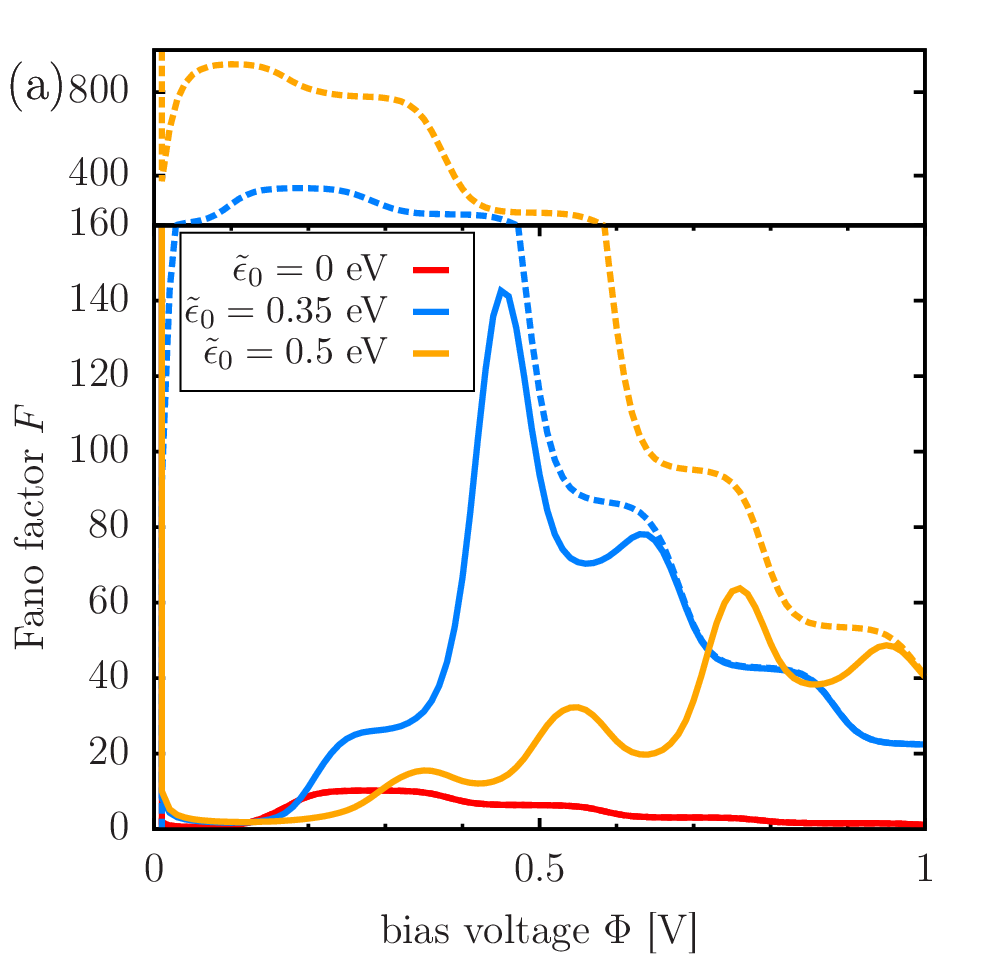} & 
	  \includegraphics[width=0.49\columnwidth]{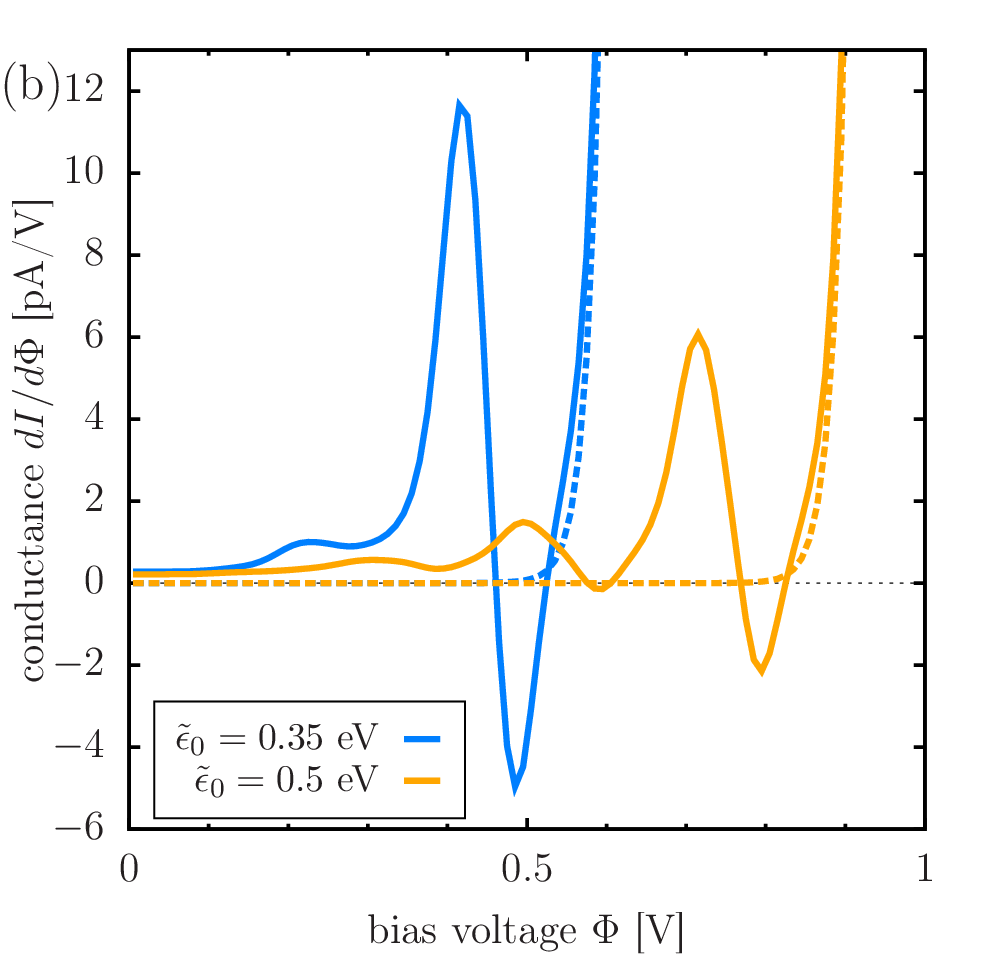}\\
	  \includegraphics[width=0.49\columnwidth]{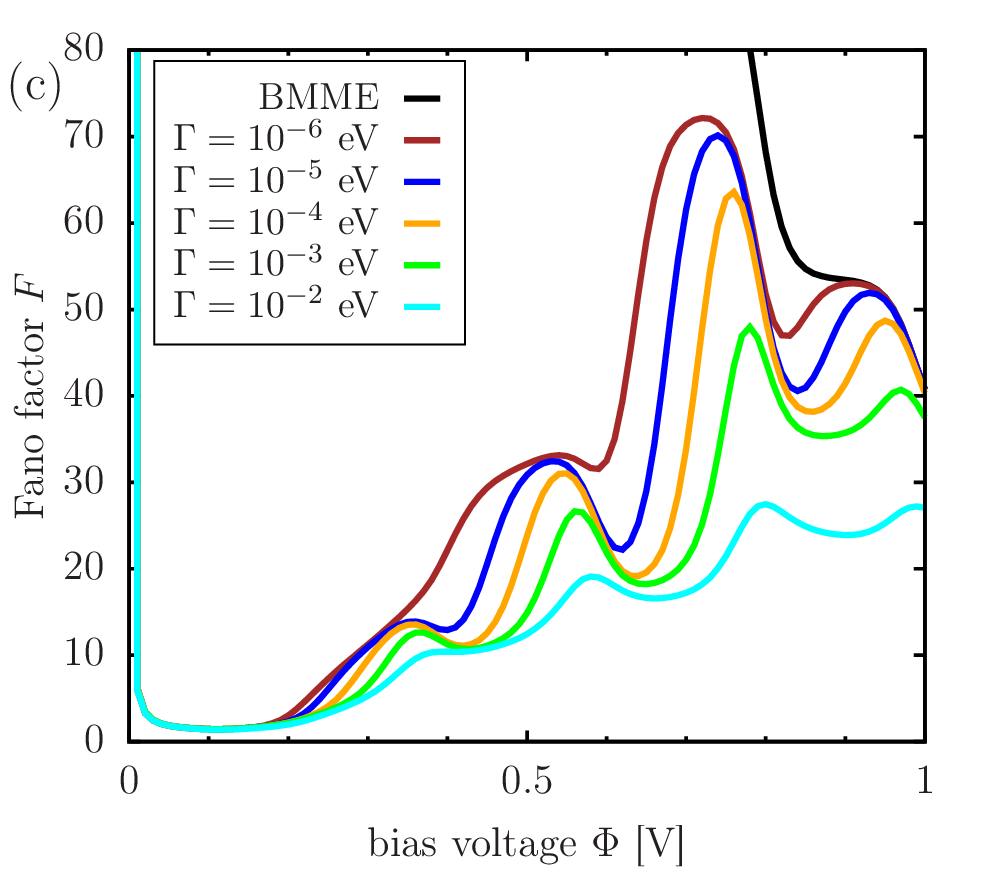} &
	  \includegraphics[width=0.49\columnwidth]{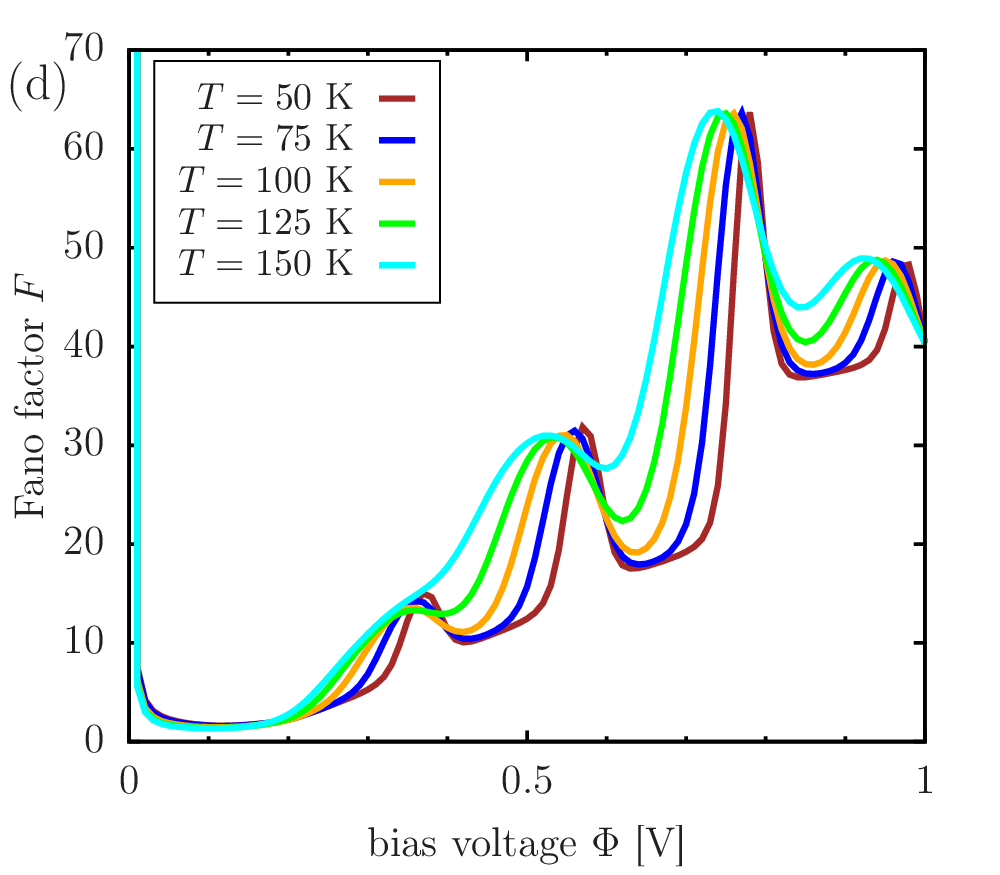}\\
	\end{tabular}
\caption{Fano factor (a,c,d) and conductance (b) as a function of voltage for $\lambda/\Omega=3$ and $\Omega=0.1 \unit{eV}$. In panels a and b, the results are shown for $T=100 \unit{K}$, $\Gamma=10^{-4} \unit{eV}$ and different energies of the molecular level. In contrast, panel c (d) depicts the Fano factor for $\tilde \epsilon_0=0.5 \unit{eV}$ and $T=100 \unit{K}$ ($\Gamma=10^{-4} \unit{eV}$) where the molecule-lead coupling $\Gamma$ (lead temperature $T$) is varied. Solid lines represent the numerically exact HQME results while dashed lines show approximate BMME results.}
\label{fig:Noise_eps}
\end{figure}

Fig.\ \ref{fig:Noise_eps}a depicts the Fano factor-voltage characteristics for different energies of the renormalized molecular level, $\tilde \epsilon_0 = \epsilon_0 -\lambda^2/\Omega$, and a set of parameters typical for molecular junctions weakly coupled to leads, $\lambda/\Omega=3$, $\Omega=0.1 \unit{eV}$, $T=100 \unit{K}$ as well as $\Gamma \equiv \Gamma_{\tL,11}=\Gamma_{\tR,11}=10^{-4} \unit{eV}$.
In the nonresonant transport regime ($\Phi < 2 \tilde \epsilon_0$), the Fano factor-voltage characteristics obtained by the HQME approach increase with bias voltage until they assume their maximum around $e \Phi \approx 2 (\tilde \epsilon_0 - \Omega)$. For higher bias voltages, they exhibit a stepwise decrease. 

\begin{figure}
	\centering
\begin{tabular}{llllllll}
(a)\hspace{-0.4cm} & &\hspace{-0.3cm} (b)\hspace{-0.4cm} & &\hspace{-0.3cm} (c)\hspace{-0.4cm} & & \hspace{-0.3cm} (d)\hspace{-0.4cm} &\\
& 
\includegraphics[width=.22\columnwidth]{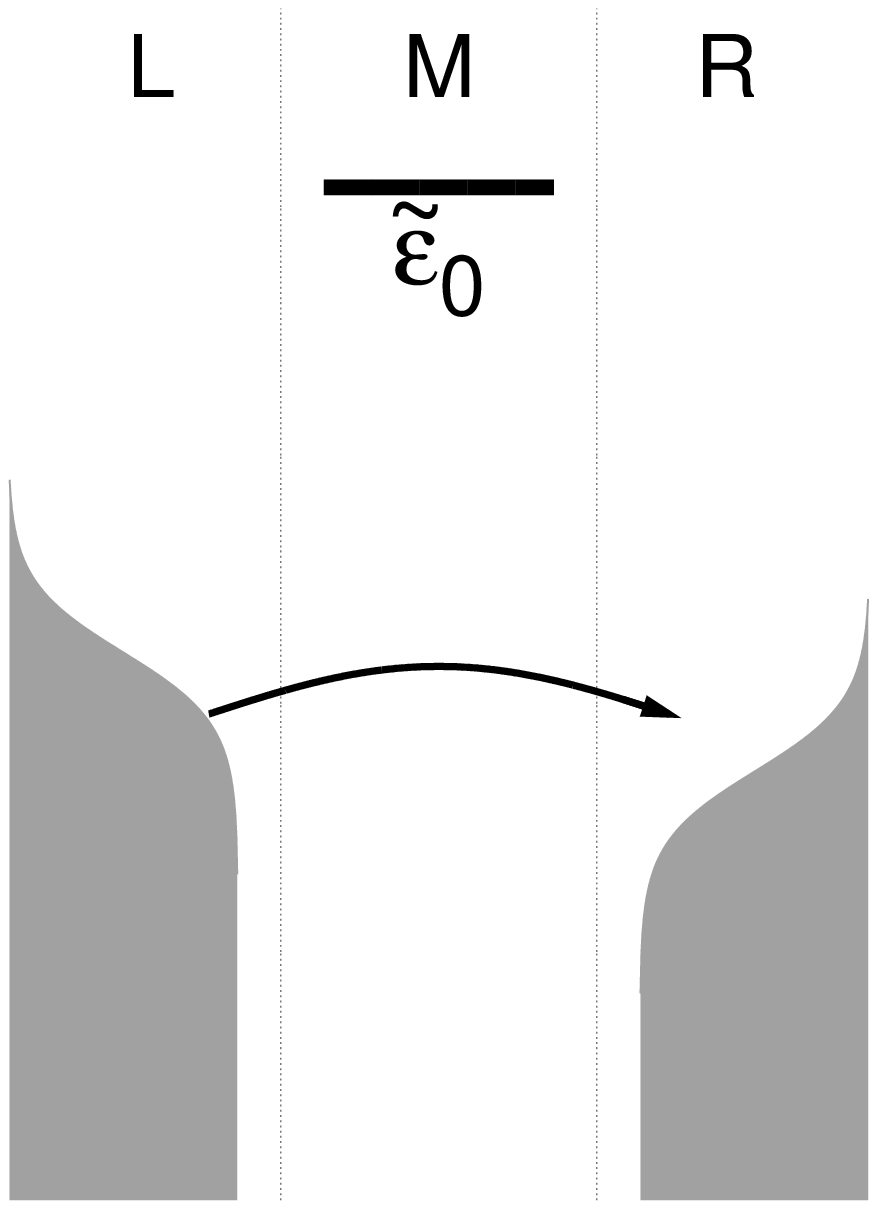}
&  &
\includegraphics[width=.22\columnwidth]{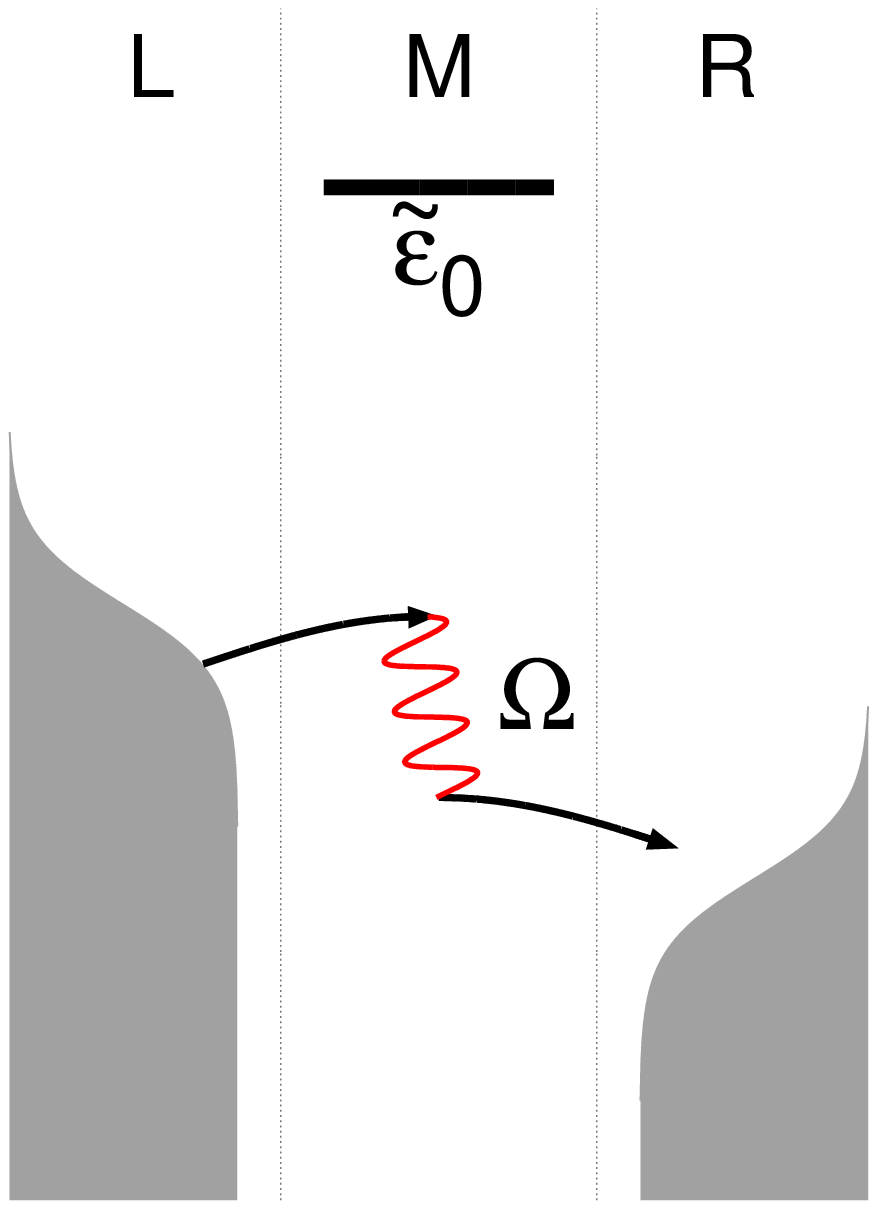}
& &
\includegraphics[width=.22\columnwidth]{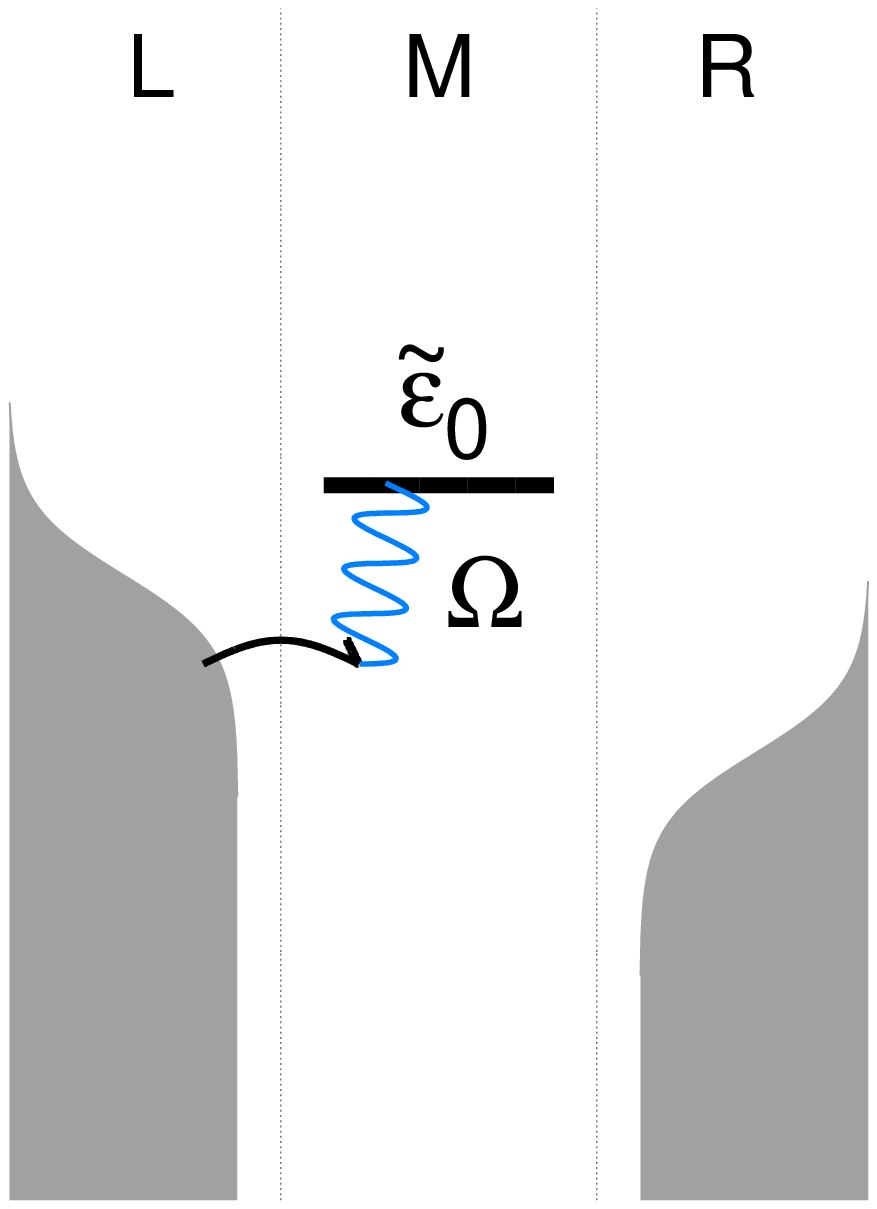}
& &
\includegraphics[width=.22\columnwidth]{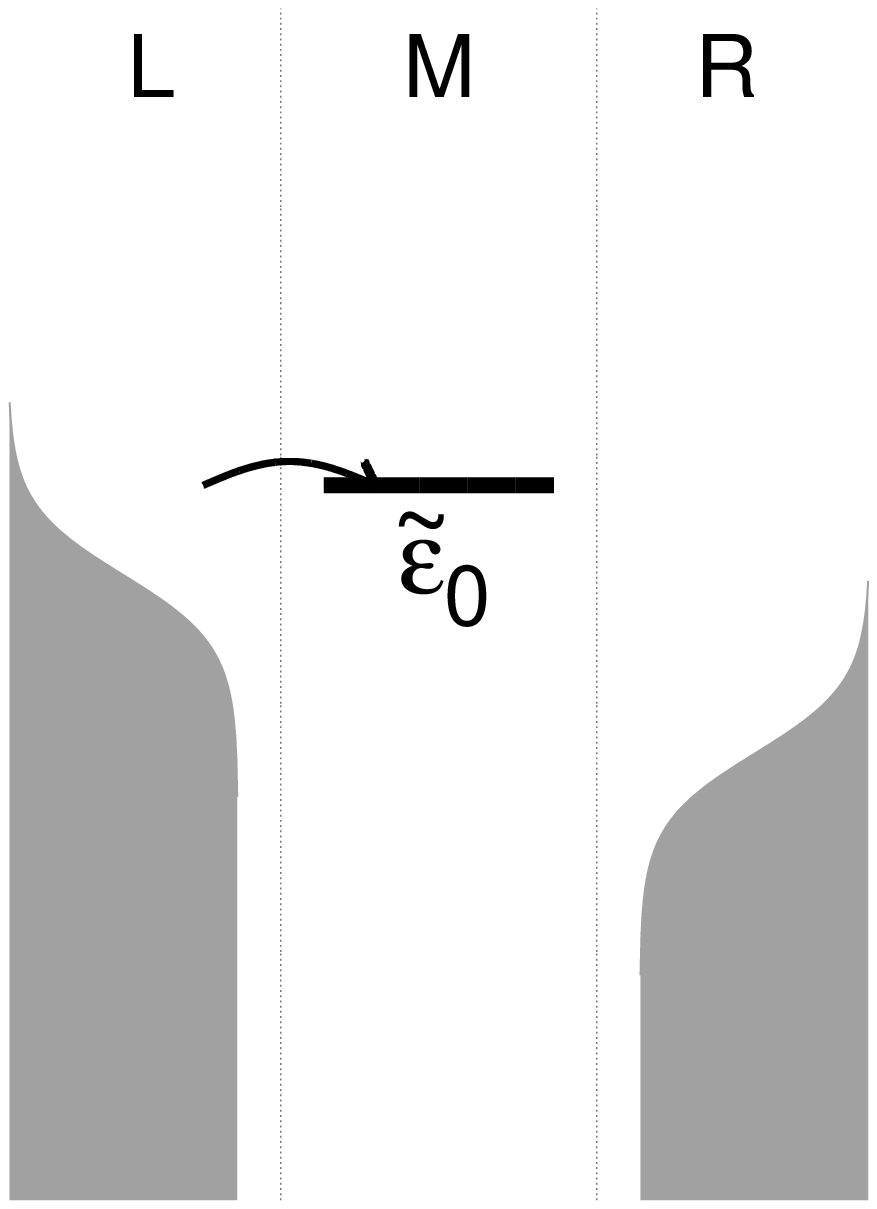}
\end{tabular}

		\caption{Illustration of transport processes in the nonresonant transport regime at nonzero temperature of the leads. In panels (a) and (b), an electron tunnels from the left to the right lead via virtual occupation of the molecular bridge. While panel (a) shows an elastic cotunneling process, a similar inelastic cotunneling process is depicted in panel (b) where the vibrational mode is excited by a single vibrational quantum of frequency $\Omega$ (red wiggly line). In panels (c) and (d), an electron sequentially tunnels from the left lead onto the molecular bridge, where the process in panel (c) includes the deexcitation of the vibrational mode by a single vibrational quantum and the process in panel (d) is enabled by the thermal broadening of the Fermi distribution.}
\label{fig:sample_processes}
\end{figure}

In the resonant transport regime ($\Phi > 2 \tilde \epsilon_0$), the HQME results are reproduced by Born-Markov (BM) calculations, which neglect higher-order cotunneling effects. In a BM picture, electron transport takes place in resonant avalanches with long waiting times in between. The detailed structure of these avalanches, which start and terminate in the vibrational ground state, has been discussed in Refs.\ \onlinecite{Koch2006,Schinabeck2014}. The corresponding  Fano factor is determined by the average number of electrons in a such an avalanche \cite{Koch2005FCS,Schinabeck2014}.
In contrast, in the nonresonant transport regime, the accurate HQME approach predicts significantly smaller Fano factors compared to the BM method because of the coexistence of resonant avalanches and elastic or inelastic cotunneling (cf.\ Fig.\ \ref{fig:sample_processes}a,b). The latter processes, which lead to Fano factors of the order of one, govern the waiting times \cite{Koch2006}, which are considerably longer than in the resonant transport regime because both chemical potentials are below the Fermi energy.

For $\tilde \epsilon_0=0.5 \unit{eV}$, the Fano factor-voltage characteristics exhibits an oscillatory behavior, which is not captured by the BM approach. While the bias voltages where the Fano factor drops agree with the stepwise decrease in the BM calculation, the bias voltages where this observable increases correspond to peaks in the differential conductance depicted in Fig.\ \ref{fig:Noise_eps}b. These peaks represent a signature of resonant absorption processes (cf.\ Fig.\ \ref{fig:sample_processes}c) facilitated by the nonequilibrium vibrational excitation induced by inelastic cotunneling.
The combination of these processes provides a trigger process for resonant avalanches in the nonresonant transport regime in addition to the sequential process depicted in Fig.\ \ref{fig:sample_processes}d, which has to be facilitated by thermal broadening.
Figs.\ \ref{fig:Noise_eps}c and \ref{fig:Noise_eps}d, which depict the Fano factor-voltage characteristics for different lead temperatures $T$ and different molecule-lead coupling strengths $\Gamma$, demonstrate that the resonance corresponding to the onset of these trigger processes is shifted to lower bias voltages with increasing lead temperature $T$ and decreasing molecule-lead coupling $\Gamma$.
This behavior can be rationalized as follows. The resonance requires a significant change in the populations \cite{Golovach2004,Gergs2015}, i.e.\ the probability for deexcitation by resonant absorption has to become higher than for excitation by inelastic cotunneling. The different dependence of these processes on temperature and molecule-lead coupling strength explains the observed behavior.
A closer inspection of Fig.\ \ref{fig:Noise_eps}d reveals that the width of the peaks in the Fano factor, that is the distance between the increase and the subsequent decrease of the Fano factor, is proportional to temperature. Consequently, we can conclude that the oscillations are a finite temperature effect, whereas the Fano factor is supposed to exhibit a stepwise increase for zero temperature in the nonresonant transport regime.

\begin{figure}[b!]
\centering
\includegraphics[width=1.0\columnwidth]{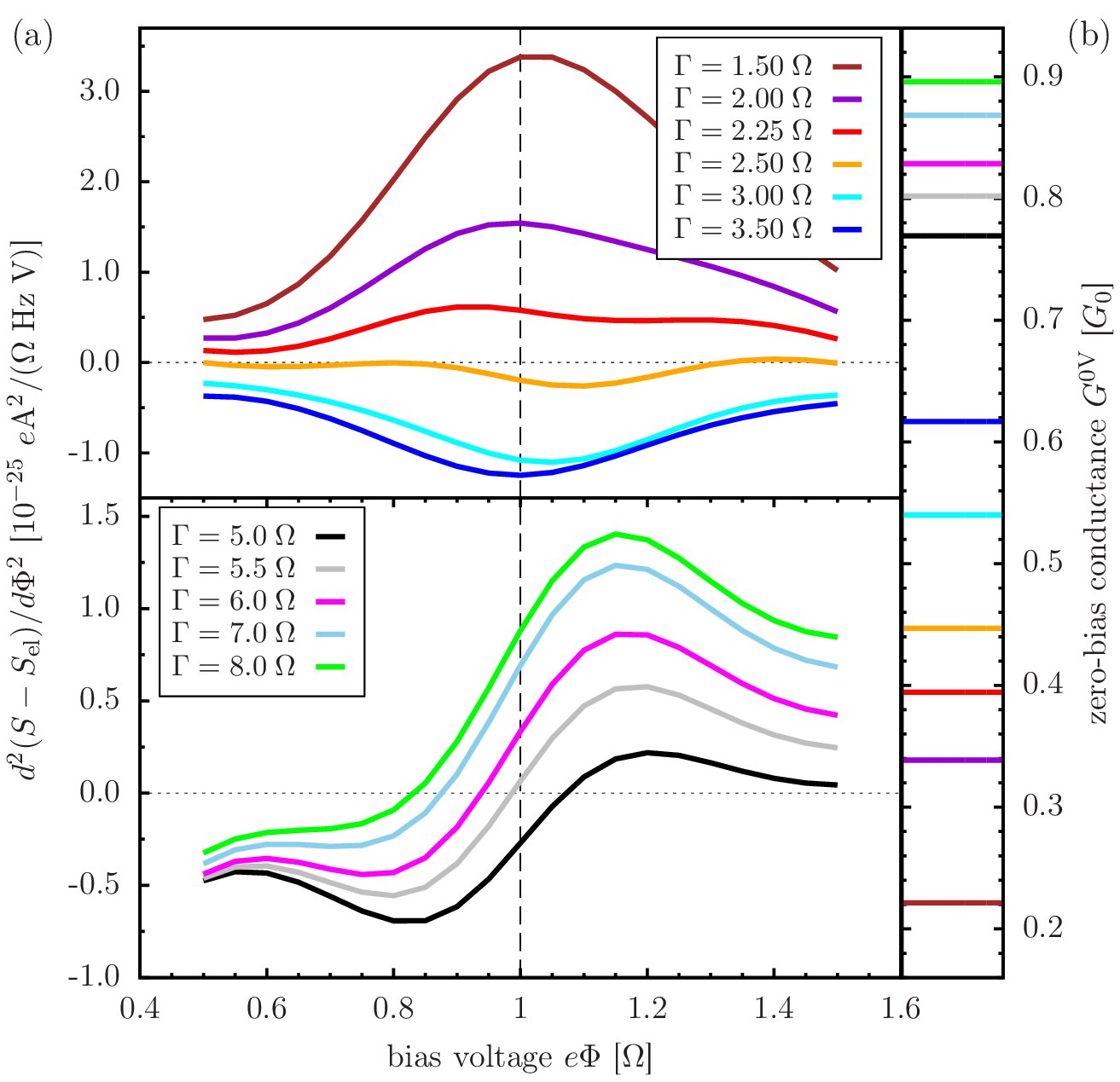}
\caption{Inelastic correction to noise (a) measured by the second derivative of the noise with respect to bias voltage for $\epsilon_0=3\, \Omega$, $\lambda/\Omega=0.6$, $e\Phi=\Omega$, $k_\tB T=0.0862\, \Omega$ and different values of the molecule-lead coupling strength $\Gamma$.
For a better resolution of inelastic effects, the signal $S_\text{el}$ of a noninteracting single-level model with level energy $\epsilon_0$ is subtracted from the total noise in order to approximately remove the elastic background. 
The corresponding zero-bias conductance is depicted in panel b.
All results have been obtained using a truncation of the hierarchy at the fourth level. 
}
\label{fig:IETS_noise_gamma}
\end{figure}
Another important aspect of vibrationally-coupled electron transport in nanosystems has been to understand how inelastic effects alter the transport characteristics, in particular close to the threshold bias voltage $e \Phi=\Omega$ \cite{Avriller2009,Schmidt2009,Haupt2010,Novotny2011,Kumar2012,Avriller2012,Utsumi2013}.
In analogy to the current, the onset of inelastic cotunneling at $e \Phi=\Omega$ leads to a negative or positive contribution to the noise, which is reflected by a positive or negative sign of the second derivative of noise with respect to bias voltage.
As a representative example, Fig.\ \ref{fig:IETS_noise_gamma}a depicts this observable for parameters $\epsilon_0=3\, \Omega$, $\lambda/\Omega=0.6$, $k_\tB T=0.0862\, \Omega$.
Considering the results obtained by a fourth-tier truncation of the HQMEs, we find two transitions: The first one from a positive to a negative contribution between $\Gamma=2.25\, \Omega$ and $\Gamma=2.5\, \Omega$ corresponding to a zero-bias conductance $G^\text{0V}$ of 0.395 and 0.445 $G_0$, respectively (cf.\ Fig.\ \ref{fig:IETS_noise_gamma}b), where $G_0$ denotes the conductance quantum.
The second crossover from a negative to a positive sign occurs around $\Gamma=5.5\, \Omega$ ($0.8\, G_0$).
These transitions can be rationalized by the competition between one-electron and coherent two-electron processes, where the former (latter) always gives a positive (negative) contribution to the noise \cite{Kumar2012}.
%

\begin{figure}[h!]
	\centering
\includegraphics[width=.9\columnwidth]{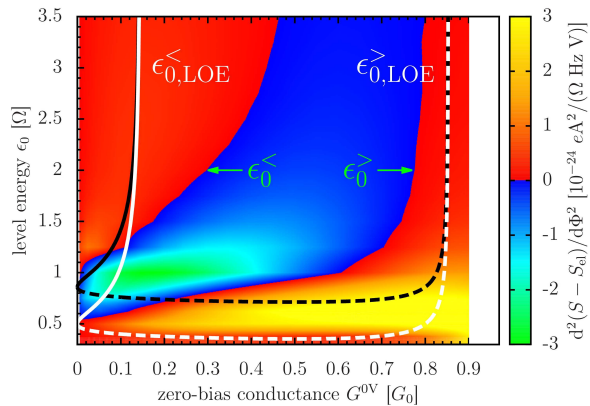}\\
		\caption{Inelastic correction to noise at $e\Phi=\Omega$ as a function of zero-bias conductance $G^\text{0V}$ and energy $\epsilon_0$ of the molecular level. The color map  was obtained by a fourth-tier truncation of the HQME approach for $\lambda/\Omega=0.6$ and $k_\tB T=0.0862\, \Omega$. 
		For comparison, the solid [dashed] white line shows the transition branch $\epsilon_{0,\text{LOE}}^< (G^\text{0V})$ $[\epsilon_{0,\text{LOE}}^> (G^\text{0V})]$ from positive to negative [negative to positive] inelastic correction obtained on the basis of a LOE in electronic-vibrational coupling, which includes the full energy dependence of the model but treats the vibrational mode in thermal equilibrium \cite{Avriller2009}.
Taking into account the polaron shift, the prediction of the LOE changes to the black lines, $\epsilon^{</>}_{0,\text{LOE}} (G^\text{0V}) \to \epsilon^{</>}_{0,\text{LOE}} (G^\text{0V}) + \lambda^2/\Omega$.}
\label{fig:Noise_eps_LOE}
\end{figure}
It is important to emphasize, that the crossover does not only depend on the zero-bias conductance but, in principle, on all model parameters.
This is demonstrated in the color map in Fig.\ \ref{fig:Noise_eps_LOE}, which depicts the inelastic correction to the noise as a function of zero-bias conductance and energy of the molecular level obtained within the HQME approach.
To analyze the influence of the nonequilibrium vibrational excitation, the white lines mark the transitions between positive and negative inelastic corrections as obtained by a lowest-order expansion (LOE) in electronic-vibrational coupling, where the full energy dependence of the model is taken into account but the vibration is treated in thermal equilibrium \cite{Avriller2009}.
%
There are two transition branches $\epsilon_{0,\mathrm{LOE}}^{</>}(G^\text{0V})$ (indicated by solid/dashed white lines).
In the limit $\epsilon_0 \gg \Omega$, the crossovers occur at $(1/2 \pm 1/\sqrt{2}) G_0$ and thus follow the predictions \cite{Avriller2009,Schmidt2009,Kumar2012} obtained within the extended wide-band approximation \cite{Haupt2010,Avriller2012}.
While there is only one transition from negative to positive sign for $\epsilon_0 =0.5\, \Omega$, the inelastic correction is always positive for $\epsilon_0 \leq \Omega/(2 \sqrt{2})$.
Comparing these LOE results with the two transition branches 
obtained by the HQME calculations (marked as $\epsilon_{0}^{</>}$ in Fig.\ \ref{fig:Noise_eps_LOE}), we find that the transitions are shifted to $0.46\, G_0$ and $0.80\, G_0$ for $\epsilon_0 =3.5\, \Omega$. This is a consequence of the nonequilibrium vibrational excitation.
In contrast to the upper transition, the lower transition is not constant for $\epsilon_0 > 3\, \Omega$ yet but still increases which demonstrates that the limit $\epsilon_0 \gg \Omega$ where the extended wide-band approximation applies is not yet reached for $\epsilon_0 =3.5\, \Omega$ in the nonequilibrium case.
In this limit, perturbative calculations of Novotny \emph{et al.} \cite{Novotny2011} and Utsumi \emph{et al.} \cite{Utsumi2013} predict the crossovers around a zero-bias conductance of $0.434\, G_0$ and $0.816\, G_0$.
In these studies, the nonequilibrium vibrational distribution function was systematically included in the LOE description by a self-consistent scheme based on the Luttinger-Ward functional \cite{Utsumi2013}
or by an infinite resummation of the electron-hole polarization bubble \cite{Viljas2005,Haupt2010,Novotny2011}.

Next, we investigate how the transitions are influenced by the energy renormalization of the molecular level due to electronic-vibrational coupling, $\epsilon_0 \to \epsilon_0 - \lambda^2/\Omega$ (polaron shift). This effect is not included in a LOE description. 
In the limit of vanishing zero-bias conductance, the branch $\epsilon_0^> (G^\text{0V})$ follows the prediction of the LOE for an equilibrated vibration, whereas the branch $\epsilon_0^< (G^\text{0V}) $ agrees reasonably well with $\epsilon_{0,\text{LOE}}^< (G^\text{0V}) +\lambda^2/\Omega$, which demonstrates that the polaron shift is completely developed in this regime. Consequently, a gap opens for $\epsilon_0 \in [\Omega/2,\Omega]$ where only one transition occurs.
The branch $\epsilon_0^> (G^\text{0V})$ increases with rising zero-bias conductance and even crosses $\epsilon_{0,\text{LOE}}^>(G^\text{0V}) +\lambda^2/\Omega$ (dashed black line) for $G^\text{0V} \approx G_0/4$.
This behavior might also be attributed to the polaron shift which seems to be fully developed for higher conductance values, whereas it does not play a role for $G^\text{0V} \to 0$.

We finally comment on the convergence of the HQME method. A comparison of different hierarchy truncation levels $n$ (data not shown) demonstrates that the HQME results for the conductance and the noise in Fig.\ \ref{fig:Noise_eps} are quantitatively converged already for $n = 2$. For the data shown in Figs.\ \ref{fig:IETS_noise_gamma} and \ref{fig:Noise_eps_LOE}, a truncation at $n=4$ was used. For very large couplings $\Gamma$, an even higher truncation would be required to obtain fully quantitative results. This is to be expected because the second derivative of the noise is significantly more difficult to converge than the noise itself or the conductance.


In summary, the HQME method presented here allows the numerically exact evaluation of current fluctuations in nonequilibrium nanosystems with strong electronic-vibrational coupling. 
The results obtained for a generic model of vibrationally coupled electron transport in molecular junctions demonstrate the importance of cotunneling effects in the nonresonant transport regime and show that the transition from positive to negative inelastic corrections to shot noise depend in a non-universal way on the physical parameters.

While in the present study we have focused on the study of the zero-frequency noise, the method is formulated in a way that it can also describe higher-order current cumulants and the full counting statistics. Furthermore, it can also be applied to models with additional electron-electron interaction \cite{Haertle2015} and extended to treat explicitly time-dependent problems \cite{Benito2016}.


We thank A.\ Erpenbeck, R. {H\"artle}, T.\ Novotny, and H.\ Weber for fruitful and inspiring discussions. This work was
supported by the German Research Foundation (DFG) via
SFB 953 and a research grant.

\bibliography{./mybib}

\begin{thebibliography}{98}%
\makeatletter
\providecommand \@ifxundefined [1]{%
 \@ifx{#1\undefined}
}%
\providecommand \@ifnum [1]{%
 \ifnum #1\expandafter \@firstoftwo
 \else \expandafter \@secondoftwo
 \fi
}%
\providecommand \@ifx [1]{%
 \ifx #1\expandafter \@firstoftwo
 \else \expandafter \@secondoftwo
 \fi
}%
\providecommand \natexlab [1]{#1}%
\providecommand \enquote  [1]{``#1''}%
\providecommand \bibnamefont  [1]{#1}%
\providecommand \bibfnamefont [1]{#1}%
\providecommand \citenamefont [1]{#1}%
\providecommand \href@noop [0]{\@secondoftwo}%
\providecommand \href [0]{\begingroup \@sanitize@url \@href}%
\providecommand \@href[1]{\@@startlink{#1}\@@href}%
\providecommand \@@href[1]{\endgroup#1\@@endlink}%
\providecommand \@sanitize@url [0]{\catcode `\\12\catcode `\$12\catcode
  `\&12\catcode `\#12\catcode `\^12\catcode `\_12\catcode `\%12\relax}%
\providecommand \@@startlink[1]{}%
\providecommand \@@endlink[0]{}%
\providecommand \url  [0]{\begingroup\@sanitize@url \@url }%
\providecommand \@url [1]{\endgroup\@href {#1}{\urlprefix }}%
\providecommand \urlprefix  [0]{URL }%
\providecommand \Eprint [0]{\href }%
\providecommand \doibase [0]{https://doi.org/}%
\providecommand \selectlanguage [0]{\@gobble}%
\providecommand \bibinfo  [0]{\@secondoftwo}%
\providecommand \bibfield  [0]{\@secondoftwo}%
\providecommand \translation [1]{[#1]}%
\providecommand \BibitemOpen [0]{}%
\providecommand \bibitemStop [0]{}%
\providecommand \bibitemNoStop [0]{.\EOS\space}%
\providecommand \EOS [0]{\spacefactor3000\relax}%
\providecommand \BibitemShut  [1]{\csname bibitem#1\endcsname}%
\let\auto@bib@innerbib\@empty
\bibitem [{\citenamefont {Levitov}\ and\ \citenamefont
  {Lesovik}(1993)}]{Levitov1993}%
  \BibitemOpen
  \bibfield  {author} {\bibinfo {author} {\bibfnamefont {L.~S.}\ \bibnamefont
  {Levitov}}\ and\ \bibinfo {author} {\bibfnamefont {G.~B.}\ \bibnamefont
  {Lesovik}},\ }\bibfield  {title} {\bibinfo {title} {Charge distribution in
  quantum shot noise},\ }\href
  {http://www.jetpletters.ac.ru/ps/1186/article_17907.shtml} {\bibfield
  {journal} {\bibinfo  {journal} {JETP Lett.}\ }\textbf {\bibinfo {volume}
  {58}},\ \bibinfo {pages} {230} (\bibinfo {year} {1993})}\BibitemShut
  {NoStop}%
\bibitem [{\citenamefont {Lee}\ \emph {et~al.}(1995)\citenamefont {Lee},
  \citenamefont {Levitov},\ and\ \citenamefont {Yakovets}}]{Lee1995}%
  \BibitemOpen
  \bibfield  {author} {\bibinfo {author} {\bibfnamefont {H.}~\bibnamefont
  {Lee}}, \bibinfo {author} {\bibfnamefont {L.~S.}\ \bibnamefont {Levitov}},\
  and\ \bibinfo {author} {\bibfnamefont {A.~Y.}\ \bibnamefont {Yakovets}},\
  }\bibfield  {title} {\bibinfo {title} {Universal statistics of transport in
  disordered conductors},\ }\href {https://doi.org/10.1103/PhysRevB.51.4079}
  {\bibfield  {journal} {\bibinfo  {journal} {Phys. Rev. B}\ }\textbf {\bibinfo
  {volume} {51}},\ \bibinfo {pages} {4079} (\bibinfo {year}
  {1995})}\BibitemShut {NoStop}%
\bibitem [{\citenamefont {Levitov}\ \emph {et~al.}(1996)\citenamefont
  {Levitov}, \citenamefont {Lee},\ and\ \citenamefont {Lesovik}}]{Levitov1996}%
  \BibitemOpen
  \bibfield  {author} {\bibinfo {author} {\bibfnamefont {L.~S.}\ \bibnamefont
  {Levitov}}, \bibinfo {author} {\bibfnamefont {H.}~\bibnamefont {Lee}},\ and\
  \bibinfo {author} {\bibfnamefont {G.~B.}\ \bibnamefont {Lesovik}},\
  }\bibfield  {title} {\bibinfo {title} {Electron counting statistics and
  coherent states of electric current},\ }\href
  {https://doi.org/10.1063/1.531672} {\bibfield  {journal} {\bibinfo  {journal}
  {J. Math. Phys.}\ }\textbf {\bibinfo {volume} {37}},\ \bibinfo {pages} {4845}
  (\bibinfo {year} {1996})}\BibitemShut {NoStop}%
\bibitem [{\citenamefont {Reulet}\ \emph {et~al.}(2003)\citenamefont {Reulet},
  \citenamefont {Senzier},\ and\ \citenamefont {Prober}}]{Reulet2003}%
  \BibitemOpen
  \bibfield  {author} {\bibinfo {author} {\bibfnamefont {B.}~\bibnamefont
  {Reulet}}, \bibinfo {author} {\bibfnamefont {J.}~\bibnamefont {Senzier}},\
  and\ \bibinfo {author} {\bibfnamefont {D.~E.}\ \bibnamefont {Prober}},\
  }\bibfield  {title} {\bibinfo {title} {Environmental effects in the third
  moment of voltage fluctuations in a tunnel junction},\ }\href
  {https://doi.org/10.1103/PhysRevLett.91.196601} {\bibfield  {journal}
  {\bibinfo  {journal} {Phys. Rev. Lett.}\ }\textbf {\bibinfo {volume} {91}},\
  \bibinfo {pages} {196601} (\bibinfo {year} {2003})}\BibitemShut {NoStop}%
\bibitem [{\citenamefont {Bomze}\ \emph {et~al.}(2005)\citenamefont {Bomze},
  \citenamefont {Gershon}, \citenamefont {Shovkun}, \citenamefont {Levitov},\
  and\ \citenamefont {Reznikov}}]{Bomze2005}%
  \BibitemOpen
  \bibfield  {author} {\bibinfo {author} {\bibfnamefont {Y.}~\bibnamefont
  {Bomze}}, \bibinfo {author} {\bibfnamefont {G.}~\bibnamefont {Gershon}},
  \bibinfo {author} {\bibfnamefont {D.}~\bibnamefont {Shovkun}}, \bibinfo
  {author} {\bibfnamefont {L.~S.}\ \bibnamefont {Levitov}},\ and\ \bibinfo
  {author} {\bibfnamefont {M.}~\bibnamefont {Reznikov}},\ }\bibfield  {title}
  {\bibinfo {title} {Measurement of counting statistics of electron transport
  in a tunnel junction},\ }\href
  {https://doi.org/10.1103/PhysRevLett.95.176601} {\bibfield  {journal}
  {\bibinfo  {journal} {Phys. Rev. Lett.}\ }\textbf {\bibinfo {volume} {95}},\
  \bibinfo {pages} {176601} (\bibinfo {year} {2005})}\BibitemShut {NoStop}%
\bibitem [{\citenamefont {Gustavsson}\ \emph {et~al.}(2006)\citenamefont
  {Gustavsson}, \citenamefont {Leturcq}, \citenamefont
  {Simovi\ifmmode~\check{c}\else \v{c}\fi{}}, \citenamefont {Schleser},
  \citenamefont {Ihn}, \citenamefont {Studerus}, \citenamefont {Ensslin},
  \citenamefont {Driscoll},\ and\ \citenamefont {Gossard}}]{Gustavsson2006}%
  \BibitemOpen
  \bibfield  {author} {\bibinfo {author} {\bibfnamefont {S.}~\bibnamefont
  {Gustavsson}}, \bibinfo {author} {\bibfnamefont {R.}~\bibnamefont {Leturcq}},
  \bibinfo {author} {\bibfnamefont {B.}~\bibnamefont
  {Simovi\ifmmode~\check{c}\else \v{c}\fi{}}}, \bibinfo {author} {\bibfnamefont
  {R.}~\bibnamefont {Schleser}}, \bibinfo {author} {\bibfnamefont
  {T.}~\bibnamefont {Ihn}}, \bibinfo {author} {\bibfnamefont {P.}~\bibnamefont
  {Studerus}}, \bibinfo {author} {\bibfnamefont {K.}~\bibnamefont {Ensslin}},
  \bibinfo {author} {\bibfnamefont {D.~C.}\ \bibnamefont {Driscoll}},\ and\
  \bibinfo {author} {\bibfnamefont {A.~C.}\ \bibnamefont {Gossard}},\
  }\bibfield  {title} {\bibinfo {title} {Counting statistics of single electron
  transport in a quantum dot},\ }\href
  {https://doi.org/10.1103/PhysRevLett.96.076605} {\bibfield  {journal}
  {\bibinfo  {journal} {Phys. Rev. Lett.}\ }\textbf {\bibinfo {volume} {96}},\
  \bibinfo {pages} {076605} (\bibinfo {year} {2006})}\BibitemShut {NoStop}%
\bibitem [{\citenamefont {Ubbelohde}\ \emph {et~al.}(2012)\citenamefont
  {Ubbelohde}, \citenamefont {Fricke}, \citenamefont {Flindt}, \citenamefont
  {Hohls},\ and\ \citenamefont {Haug}}]{Ubbelohde2012}%
  \BibitemOpen
  \bibfield  {author} {\bibinfo {author} {\bibfnamefont {N.}~\bibnamefont
  {Ubbelohde}}, \bibinfo {author} {\bibfnamefont {C.}~\bibnamefont {Fricke}},
  \bibinfo {author} {\bibfnamefont {C.}~\bibnamefont {Flindt}}, \bibinfo
  {author} {\bibfnamefont {F.}~\bibnamefont {Hohls}},\ and\ \bibinfo {author}
  {\bibfnamefont {R.~J.}\ \bibnamefont {Haug}},\ }\bibfield  {title} {\bibinfo
  {title} {Measurement of finite-frequency current statistics in a
  single-electron transistor},\ }\href@noop {} {\bibfield  {journal} {\bibinfo
  {journal} {Nat. Commun.}\ }\textbf {\bibinfo {volume} {3}},\ \bibinfo {pages}
  {612} (\bibinfo {year} {2012})}\BibitemShut {NoStop}%
\bibitem [{\citenamefont {de~Jong}\ and\ \citenamefont
  {Beenakker}(1994)}]{de_Jong1994}%
  \BibitemOpen
  \bibfield  {author} {\bibinfo {author} {\bibfnamefont {M.~J.~M.}\
  \bibnamefont {de~Jong}}\ and\ \bibinfo {author} {\bibfnamefont {C.~W.~J.}\
  \bibnamefont {Beenakker}},\ }\bibfield  {title} {\bibinfo {title} {Doubled
  shot noise in disordered normal-metal--superconductor junctions},\ }\href
  {https://doi.org/10.1103/PhysRevB.49.16070} {\bibfield  {journal} {\bibinfo
  {journal} {Phys. Rev. B}\ }\textbf {\bibinfo {volume} {49}},\ \bibinfo
  {pages} {16070} (\bibinfo {year} {1994})}\BibitemShut {NoStop}%
\bibitem [{\citenamefont {Jehl}\ \emph {et~al.}(2000)\citenamefont {Jehl},
  \citenamefont {Sanquer}, \citenamefont {Calemczuk},\ and\ \citenamefont
  {Mailly}}]{Jehl2000}%
  \BibitemOpen
  \bibfield  {author} {\bibinfo {author} {\bibfnamefont {X.}~\bibnamefont
  {Jehl}}, \bibinfo {author} {\bibfnamefont {M.}~\bibnamefont {Sanquer}},
  \bibinfo {author} {\bibfnamefont {R.}~\bibnamefont {Calemczuk}},\ and\
  \bibinfo {author} {\bibfnamefont {D.}~\bibnamefont {Mailly}},\ }\bibfield
  {title} {\bibinfo {title} {Detection of doubled shot noise in short
  normal-metal/ superconductor junctions},\ }\href
  {https://doi.org/10.1038/35011012} {\bibfield  {journal} {\bibinfo  {journal}
  {Nature}\ }\textbf {\bibinfo {volume} {405}},\ \bibinfo {pages} {50}
  (\bibinfo {year} {2000})}\BibitemShut {NoStop}%
\bibitem [{\citenamefont {Lefloch}\ \emph {et~al.}(2003)\citenamefont
  {Lefloch}, \citenamefont {Hoffmann}, \citenamefont {Sanquer},\ and\
  \citenamefont {Quirion}}]{Lefloch2003}%
  \BibitemOpen
  \bibfield  {author} {\bibinfo {author} {\bibfnamefont {F.}~\bibnamefont
  {Lefloch}}, \bibinfo {author} {\bibfnamefont {C.}~\bibnamefont {Hoffmann}},
  \bibinfo {author} {\bibfnamefont {M.}~\bibnamefont {Sanquer}},\ and\ \bibinfo
  {author} {\bibfnamefont {D.}~\bibnamefont {Quirion}},\ }\bibfield  {title}
  {\bibinfo {title} {Doubled full shot noise in quantum coherent
  superconductor-semiconductor junctions},\ }\href
  {https://doi.org/10.1103/PhysRevLett.90.067002} {\bibfield  {journal}
  {\bibinfo  {journal} {Phys. Rev. Lett.}\ }\textbf {\bibinfo {volume} {90}},\
  \bibinfo {pages} {067002} (\bibinfo {year} {2003})}\BibitemShut {NoStop}%
\bibitem [{\citenamefont {de~Picciotto}\ \emph {et~al.}(1997)\citenamefont
  {de~Picciotto}, \citenamefont {Reznikov}, \citenamefont {Heiblum},
  \citenamefont {Umansky}, \citenamefont {Bunin},\ and\ \citenamefont
  {Mahalu}}]{Picciotto1997}%
  \BibitemOpen
  \bibfield  {author} {\bibinfo {author} {\bibfnamefont {R.}~\bibnamefont
  {de~Picciotto}}, \bibinfo {author} {\bibfnamefont {M.}~\bibnamefont
  {Reznikov}}, \bibinfo {author} {\bibfnamefont {M.}~\bibnamefont {Heiblum}},
  \bibinfo {author} {\bibfnamefont {V.}~\bibnamefont {Umansky}}, \bibinfo
  {author} {\bibfnamefont {G.}~\bibnamefont {Bunin}},\ and\ \bibinfo {author}
  {\bibfnamefont {D.}~\bibnamefont {Mahalu}},\ }\bibfield  {title} {\bibinfo
  {title} {Direct observation of a fractional charge},\ }\href@noop {}
  {\bibfield  {journal} {\bibinfo  {journal} {Nature}\ }\textbf {\bibinfo
  {volume} {389}},\ \bibinfo {pages} {162} (\bibinfo {year}
  {1997})}\BibitemShut {NoStop}%
\bibitem [{\citenamefont {Saminadayar}\ \emph {et~al.}(1997)\citenamefont
  {Saminadayar}, \citenamefont {Glattli}, \citenamefont {Jin},\ and\
  \citenamefont {Etienne}}]{Saminadayar1997}%
  \BibitemOpen
  \bibfield  {author} {\bibinfo {author} {\bibfnamefont {L.}~\bibnamefont
  {Saminadayar}}, \bibinfo {author} {\bibfnamefont {D.~C.}\ \bibnamefont
  {Glattli}}, \bibinfo {author} {\bibfnamefont {Y.}~\bibnamefont {Jin}},\ and\
  \bibinfo {author} {\bibfnamefont {B.}~\bibnamefont {Etienne}},\ }\bibfield
  {title} {\bibinfo {title} {Observation of the $\mathit{e}\mathit{/}3$
  fractionally charged laughlin quasiparticle},\ }\href
  {https://doi.org/10.1103/PhysRevLett.79.2526} {\bibfield  {journal} {\bibinfo
   {journal} {Phys. Rev. Lett.}\ }\textbf {\bibinfo {volume} {79}},\ \bibinfo
  {pages} {2526} (\bibinfo {year} {1997})}\BibitemShut {NoStop}%
\bibitem [{\citenamefont {van~den Brom}\ and\ \citenamefont {van
  Ruitenbeek}(1999)}]{Brom1999}%
  \BibitemOpen
  \bibfield  {author} {\bibinfo {author} {\bibfnamefont {H.~E.}\ \bibnamefont
  {van~den Brom}}\ and\ \bibinfo {author} {\bibfnamefont {J.~M.}\ \bibnamefont
  {van Ruitenbeek}},\ }\bibfield  {title} {\bibinfo {title} {Quantum
  suppression of shot noise in atom-size metallic contacts},\ }\href
  {https://doi.org/10.1103/PhysRevLett.82.1526} {\bibfield  {journal} {\bibinfo
   {journal} {Phys. Rev. Lett.}\ }\textbf {\bibinfo {volume} {82}},\ \bibinfo
  {pages} {1526} (\bibinfo {year} {1999})}\BibitemShut {NoStop}%
\bibitem [{\citenamefont {Cron}\ \emph {et~al.}(2001)\citenamefont {Cron},
  \citenamefont {Goffman}, \citenamefont {Esteve},\ and\ \citenamefont
  {Urbina}}]{Cron2001}%
  \BibitemOpen
  \bibfield  {author} {\bibinfo {author} {\bibfnamefont {R.}~\bibnamefont
  {Cron}}, \bibinfo {author} {\bibfnamefont {M.~F.}\ \bibnamefont {Goffman}},
  \bibinfo {author} {\bibfnamefont {D.}~\bibnamefont {Esteve}},\ and\ \bibinfo
  {author} {\bibfnamefont {C.}~\bibnamefont {Urbina}},\ }\bibfield  {title}
  {\bibinfo {title} {Multiple-charge-quanta shot noise in superconducting
  atomic contacts},\ }\href {https://doi.org/10.1103/PhysRevLett.86.4104}
  {\bibfield  {journal} {\bibinfo  {journal} {Phys. Rev. Lett.}\ }\textbf
  {\bibinfo {volume} {86}},\ \bibinfo {pages} {4104} (\bibinfo {year}
  {2001})}\BibitemShut {NoStop}%
\bibitem [{\citenamefont {Djukic}\ and\ \citenamefont {van
  Ruitenbeek}(2006)}]{Djukic2006}%
  \BibitemOpen
  \bibfield  {author} {\bibinfo {author} {\bibfnamefont {D.}~\bibnamefont
  {Djukic}}\ and\ \bibinfo {author} {\bibfnamefont {J.~M.}\ \bibnamefont {van
  Ruitenbeek}},\ }\bibfield  {title} {\bibinfo {title} {Shot noise measurements
  on a single molecule},\ }\href {https://doi.org/10.1021/nl060116e} {\bibfield
   {journal} {\bibinfo  {journal} {Nano Lett.}\ }\textbf {\bibinfo {volume}
  {6}},\ \bibinfo {pages} {789} (\bibinfo {year} {2006})}\BibitemShut {NoStop}%
\bibitem [{\citenamefont {Kiguchi}\ \emph {et~al.}(2008)\citenamefont
  {Kiguchi}, \citenamefont {Tal}, \citenamefont {Wohlthat}, \citenamefont
  {Pauly}, \citenamefont {Krieger}, \citenamefont {Djukic}, \citenamefont
  {Cuevas},\ and\ \citenamefont {van Ruitenbeek}}]{Kiguchi2008}%
  \BibitemOpen
  \bibfield  {author} {\bibinfo {author} {\bibfnamefont {M.}~\bibnamefont
  {Kiguchi}}, \bibinfo {author} {\bibfnamefont {O.}~\bibnamefont {Tal}},
  \bibinfo {author} {\bibfnamefont {S.}~\bibnamefont {Wohlthat}}, \bibinfo
  {author} {\bibfnamefont {F.}~\bibnamefont {Pauly}}, \bibinfo {author}
  {\bibfnamefont {M.}~\bibnamefont {Krieger}}, \bibinfo {author} {\bibfnamefont
  {D.}~\bibnamefont {Djukic}}, \bibinfo {author} {\bibfnamefont {J.~C.}\
  \bibnamefont {Cuevas}},\ and\ \bibinfo {author} {\bibfnamefont {J.~M.}\
  \bibnamefont {van Ruitenbeek}},\ }\bibfield  {title} {\bibinfo {title}
  {Highly conductive molecular junctions based on direct binding of benzene to
  platinum electrodes},\ }\href
  {https://doi.org/10.1103/PhysRevLett.101.046801} {\bibfield  {journal}
  {\bibinfo  {journal} {Phys. Rev. Lett.}\ }\textbf {\bibinfo {volume} {101}},\
  \bibinfo {pages} {046801} (\bibinfo {year} {2008})}\BibitemShut {NoStop}%
\bibitem [{\citenamefont {Tal}\ \emph {et~al.}(2008)\citenamefont {Tal},
  \citenamefont {Krieger}, \citenamefont {Leerink},\ and\ \citenamefont {van
  Ruitenbeek}}]{Tal2008}%
  \BibitemOpen
  \bibfield  {author} {\bibinfo {author} {\bibfnamefont {O.}~\bibnamefont
  {Tal}}, \bibinfo {author} {\bibfnamefont {M.}~\bibnamefont {Krieger}},
  \bibinfo {author} {\bibfnamefont {B.}~\bibnamefont {Leerink}},\ and\ \bibinfo
  {author} {\bibfnamefont {J.~M.}\ \bibnamefont {van Ruitenbeek}},\ }\bibfield
  {title} {\bibinfo {title} {Electron-vibration interaction in single-molecule
  junctions: From contact to tunneling regimes},\ }\href
  {https://doi.org/10.1103/PhysRevLett.100.196804} {\bibfield  {journal}
  {\bibinfo  {journal} {Phys. Rev. Lett.}\ }\textbf {\bibinfo {volume} {100}},\
  \bibinfo {pages} {196804} (\bibinfo {year} {2008})}\BibitemShut {NoStop}%
\bibitem [{\citenamefont {Wheeler}\ \emph {et~al.}(2010)\citenamefont
  {Wheeler}, \citenamefont {Russom}, \citenamefont {Evans}, \citenamefont
  {King},\ and\ \citenamefont {Natelson}}]{Wheeler2010}%
  \BibitemOpen
  \bibfield  {author} {\bibinfo {author} {\bibfnamefont {P.~J.}\ \bibnamefont
  {Wheeler}}, \bibinfo {author} {\bibfnamefont {J.~N.}\ \bibnamefont {Russom}},
  \bibinfo {author} {\bibfnamefont {K.}~\bibnamefont {Evans}}, \bibinfo
  {author} {\bibfnamefont {N.~S.}\ \bibnamefont {King}},\ and\ \bibinfo
  {author} {\bibfnamefont {D.}~\bibnamefont {Natelson}},\ }\bibfield  {title}
  {\bibinfo {title} {Shot noise suppression at room temperature in atomic-scale
  au junctions},\ }\href {https://doi.org/10.1021/nl904052r} {\bibfield
  {journal} {\bibinfo  {journal} {Nano Lett.}\ }\textbf {\bibinfo {volume}
  {10}},\ \bibinfo {pages} {1287} (\bibinfo {year} {2010})}\BibitemShut
  {NoStop}%
\bibitem [{\citenamefont {Schneider}\ \emph {et~al.}(2010)\citenamefont
  {Schneider}, \citenamefont {Schull},\ and\ \citenamefont
  {Berndt}}]{Schneider2010}%
  \BibitemOpen
  \bibfield  {author} {\bibinfo {author} {\bibfnamefont {N.~L.}\ \bibnamefont
  {Schneider}}, \bibinfo {author} {\bibfnamefont {G.}~\bibnamefont {Schull}},\
  and\ \bibinfo {author} {\bibfnamefont {R.}~\bibnamefont {Berndt}},\
  }\bibfield  {title} {\bibinfo {title} {Optical probe of quantum shot-noise
  reduction at a single-atom contact},\ }\href
  {https://doi.org/10.1103/PhysRevLett.105.026601} {\bibfield  {journal}
  {\bibinfo  {journal} {Phys. Rev. Lett.}\ }\textbf {\bibinfo {volume} {105}},\
  \bibinfo {pages} {026601} (\bibinfo {year} {2010})}\BibitemShut {NoStop}%
\bibitem [{\citenamefont {Chen}\ \emph {et~al.}(2012)\citenamefont {Chen},
  \citenamefont {Wheeler},\ and\ \citenamefont {Natelson}}]{Chen2012}%
  \BibitemOpen
  \bibfield  {author} {\bibinfo {author} {\bibfnamefont {R.}~\bibnamefont
  {Chen}}, \bibinfo {author} {\bibfnamefont {P.~J.}\ \bibnamefont {Wheeler}},\
  and\ \bibinfo {author} {\bibfnamefont {D.}~\bibnamefont {Natelson}},\
  }\bibfield  {title} {\bibinfo {title} {Excess noise in stm-style break
  junctions at room temperature},\ }\href
  {https://doi.org/10.1103/PhysRevB.85.235455} {\bibfield  {journal} {\bibinfo
  {journal} {Phys. Rev. B}\ }\textbf {\bibinfo {volume} {85}},\ \bibinfo
  {pages} {235455} (\bibinfo {year} {2012})}\BibitemShut {NoStop}%
\bibitem [{\citenamefont {Nazarov}\ and\ \citenamefont
  {Blanter}(2009)}]{Nazarov2009}%
  \BibitemOpen
  \bibfield  {author} {\bibinfo {author} {\bibfnamefont {Y.}~\bibnamefont
  {Nazarov}}\ and\ \bibinfo {author} {\bibfnamefont {Y.}~\bibnamefont
  {Blanter}},\ }\href {http://books.google.de/books?id=bjmXJOFmqZIC} {\emph
  {\bibinfo {title} {Quantum Transport: Introduction to Nanoscience}}}\
  (\bibinfo  {publisher} {Cambridge University Press},\ \bibinfo {year}
  {2009})\BibitemShut {NoStop}%
\bibitem [{\citenamefont {Hershfield}\ \emph {et~al.}(1993)\citenamefont
  {Hershfield}, \citenamefont {Davies}, \citenamefont {Hyldgaard},
  \citenamefont {Stanton},\ and\ \citenamefont {Wilkins}}]{Hershfield1993}%
  \BibitemOpen
  \bibfield  {author} {\bibinfo {author} {\bibfnamefont {S.}~\bibnamefont
  {Hershfield}}, \bibinfo {author} {\bibfnamefont {J.~H.}\ \bibnamefont
  {Davies}}, \bibinfo {author} {\bibfnamefont {P.}~\bibnamefont {Hyldgaard}},
  \bibinfo {author} {\bibfnamefont {C.~J.}\ \bibnamefont {Stanton}},\ and\
  \bibinfo {author} {\bibfnamefont {J.~W.}\ \bibnamefont {Wilkins}},\
  }\bibfield  {title} {\bibinfo {title} {Zero-frequency current noise for the
  double-tunnel-junction coulomb blockade},\ }\href
  {https://doi.org/10.1103/PhysRevB.47.1967} {\bibfield  {journal} {\bibinfo
  {journal} {Phys. Rev. B}\ }\textbf {\bibinfo {volume} {47}},\ \bibinfo
  {pages} {1967} (\bibinfo {year} {1993})}\BibitemShut {NoStop}%
\bibitem [{\citenamefont {Korotkov}(1994)}]{Korotkov1994}%
  \BibitemOpen
  \bibfield  {author} {\bibinfo {author} {\bibfnamefont {A.~N.}\ \bibnamefont
  {Korotkov}},\ }\bibfield  {title} {\bibinfo {title} {Intrinsic noise of the
  single-electron transistor},\ }\href
  {https://doi.org/10.1103/PhysRevB.49.10381} {\bibfield  {journal} {\bibinfo
  {journal} {Phys. Rev. B}\ }\textbf {\bibinfo {volume} {49}},\ \bibinfo
  {pages} {10381} (\bibinfo {year} {1994})}\BibitemShut {NoStop}%
\bibitem [{\citenamefont {Koch}\ \emph
  {et~al.}(2006{\natexlab{a}})\citenamefont {Koch}, \citenamefont {von Oppen},\
  and\ \citenamefont {Andreev}}]{Koch2006}%
  \BibitemOpen
  \bibfield  {author} {\bibinfo {author} {\bibfnamefont {J.}~\bibnamefont
  {Koch}}, \bibinfo {author} {\bibfnamefont {F.}~\bibnamefont {von Oppen}},\
  and\ \bibinfo {author} {\bibfnamefont {A.~V.}\ \bibnamefont {Andreev}},\
  }\bibfield  {title} {\bibinfo {title} {Theory of the franck-condon blockade
  regime},\ }\href {https://doi.org/10.1103/PhysRevB.74.205438} {\bibfield
  {journal} {\bibinfo  {journal} {Phys. Rev. B}\ }\textbf {\bibinfo {volume}
  {74}},\ \bibinfo {pages} {205438} (\bibinfo {year}
  {2006}{\natexlab{a}})}\BibitemShut {NoStop}%
\bibitem [{\citenamefont {Haupt}\ \emph {et~al.}(2006)\citenamefont {Haupt},
  \citenamefont {Cavaliere}, \citenamefont {Fazio},\ and\ \citenamefont
  {Sassetti}}]{Haupt2006}%
  \BibitemOpen
  \bibfield  {author} {\bibinfo {author} {\bibfnamefont {F.}~\bibnamefont
  {Haupt}}, \bibinfo {author} {\bibfnamefont {F.}~\bibnamefont {Cavaliere}},
  \bibinfo {author} {\bibfnamefont {R.}~\bibnamefont {Fazio}},\ and\ \bibinfo
  {author} {\bibfnamefont {M.}~\bibnamefont {Sassetti}},\ }\bibfield  {title}
  {\bibinfo {title} {Anomalous suppression of the shot noise in a
  nanoelectromechanical system},\ }\href
  {https://doi.org/10.1103/PhysRevB.74.205328} {\bibfield  {journal} {\bibinfo
  {journal} {Phys. Rev. B}\ }\textbf {\bibinfo {volume} {74}},\ \bibinfo
  {pages} {205328} (\bibinfo {year} {2006})}\BibitemShut {NoStop}%
\bibitem [{\citenamefont {Flindt}\ \emph {et~al.}(2008)\citenamefont {Flindt},
  \citenamefont {Novotn\'y}, \citenamefont {Braggio}, \citenamefont
  {Sassetti},\ and\ \citenamefont {Jauho}}]{Flindt2008}%
  \BibitemOpen
  \bibfield  {author} {\bibinfo {author} {\bibfnamefont {C.}~\bibnamefont
  {Flindt}}, \bibinfo {author} {\bibfnamefont {T.~c.~v.}\ \bibnamefont
  {Novotn\'y}}, \bibinfo {author} {\bibfnamefont {A.}~\bibnamefont {Braggio}},
  \bibinfo {author} {\bibfnamefont {M.}~\bibnamefont {Sassetti}},\ and\
  \bibinfo {author} {\bibfnamefont {A.-P.}\ \bibnamefont {Jauho}},\ }\bibfield
  {title} {\bibinfo {title} {Counting statistics of non-markovian quantum
  stochastic processes},\ }\href
  {https://doi.org/10.1103/PhysRevLett.100.150601} {\bibfield  {journal}
  {\bibinfo  {journal} {Phys. Rev. Lett.}\ }\textbf {\bibinfo {volume} {100}},\
  \bibinfo {pages} {150601} (\bibinfo {year} {2008})}\BibitemShut {NoStop}%
\bibitem [{\citenamefont {Emary}(2009)}]{Emary2009}%
  \BibitemOpen
  \bibfield  {author} {\bibinfo {author} {\bibfnamefont {C.}~\bibnamefont
  {Emary}},\ }\bibfield  {title} {\bibinfo {title} {Counting statistics of
  cotunneling electrons},\ }\href {https://doi.org/10.1103/PhysRevB.80.235306}
  {\bibfield  {journal} {\bibinfo  {journal} {Phys. Rev. B}\ }\textbf {\bibinfo
  {volume} {80}},\ \bibinfo {pages} {235306} (\bibinfo {year}
  {2009})}\BibitemShut {NoStop}%
\bibitem [{\citenamefont {Esposito}\ and\ \citenamefont
  {Galperin}(2010)}]{Esposito2010}%
  \BibitemOpen
  \bibfield  {author} {\bibinfo {author} {\bibfnamefont {M.}~\bibnamefont
  {Esposito}}\ and\ \bibinfo {author} {\bibfnamefont {M.}~\bibnamefont
  {Galperin}},\ }\bibfield  {title} {\bibinfo {title} {Self-consistent quantum
  master equation approach to molecular transport},\ }\href
  {https://doi.org/10.1021/jp103369s} {\bibfield  {journal} {\bibinfo
  {journal} {J. Phys. Chem. C}\ }\textbf {\bibinfo {volume} {114}},\ \bibinfo
  {pages} {20362} (\bibinfo {year} {2010})}\BibitemShut {NoStop}%
\bibitem [{\citenamefont {Schinabeck}\ \emph {et~al.}(2014)\citenamefont
  {Schinabeck}, \citenamefont {H\"artle}, \citenamefont {Weber},\ and\
  \citenamefont {Thoss}}]{Schinabeck2014}%
  \BibitemOpen
  \bibfield  {author} {\bibinfo {author} {\bibfnamefont {C.}~\bibnamefont
  {Schinabeck}}, \bibinfo {author} {\bibfnamefont {R.}~\bibnamefont
  {H\"artle}}, \bibinfo {author} {\bibfnamefont {H.~B.}\ \bibnamefont
  {Weber}},\ and\ \bibinfo {author} {\bibfnamefont {M.}~\bibnamefont {Thoss}},\
  }\bibfield  {title} {\bibinfo {title} {Current noise in single-molecule
  junctions induced by electronic-vibrational coupling},\ }\href
  {https://doi.org/10.1103/PhysRevB.90.075409} {\bibfield  {journal} {\bibinfo
  {journal} {Phys. Rev. B}\ }\textbf {\bibinfo {volume} {90}},\ \bibinfo
  {pages} {075409} (\bibinfo {year} {2014})}\BibitemShut {NoStop}%
\bibitem [{\citenamefont {Kaasbjerg}\ and\ \citenamefont
  {Belzig}(2015)}]{Kaasbjerg2015}%
  \BibitemOpen
  \bibfield  {author} {\bibinfo {author} {\bibfnamefont {K.}~\bibnamefont
  {Kaasbjerg}}\ and\ \bibinfo {author} {\bibfnamefont {W.}~\bibnamefont
  {Belzig}},\ }\bibfield  {title} {\bibinfo {title} {Full counting statistics
  and shot noise of cotunneling in quantum dots and single-molecule
  transistors},\ }\href {http://dx.doi.org/10.1103/PhysRevB.91.235413}
  {\bibfield  {journal} {\bibinfo  {journal} {Phys. Rev. B}\ }\textbf {\bibinfo
  {volume} {91}} (\bibinfo {year} {2015})}\BibitemShut {NoStop}%
\bibitem [{\citenamefont {Agarwalla}\ \emph {et~al.}(2015)\citenamefont
  {Agarwalla}, \citenamefont {Jiang},\ and\ \citenamefont
  {Segal}}]{Agarwalla2015}%
  \BibitemOpen
  \bibfield  {author} {\bibinfo {author} {\bibfnamefont {B.~K.}\ \bibnamefont
  {Agarwalla}}, \bibinfo {author} {\bibfnamefont {J.-H.}\ \bibnamefont
  {Jiang}},\ and\ \bibinfo {author} {\bibfnamefont {D.}~\bibnamefont {Segal}},\
  }\bibfield  {title} {\bibinfo {title} {Full counting statistics of
  vibrationally assisted electronic conduction: Transport and fluctuations of
  thermoelectric efficiency},\ }\href
  {https://doi.org/10.1103/PhysRevB.92.245418} {\bibfield  {journal} {\bibinfo
  {journal} {Phys. Rev. B}\ }\textbf {\bibinfo {volume} {92}},\ \bibinfo
  {pages} {245418} (\bibinfo {year} {2015})}\BibitemShut {NoStop}%
\bibitem [{\citenamefont {Kosov}(2017)}]{Kosov2017}%
  \BibitemOpen
  \bibfield  {author} {\bibinfo {author} {\bibfnamefont {D.~S.}\ \bibnamefont
  {Kosov}},\ }\bibfield  {title} {\bibinfo {title} {Waiting time distribution
  for electron transport in a molecular junction with electron-vibration
  interaction},\ }\href {https://doi.org/10.1063/1.4976561} {\bibfield
  {journal} {\bibinfo  {journal} {J. Chem. Phys.}\ }\textbf {\bibinfo {volume}
  {146}},\ \bibinfo {pages} {074102} (\bibinfo {year} {2017})}\BibitemShut
  {NoStop}%
\bibitem [{\citenamefont {Rudge}\ and\ \citenamefont
  {Kosov}(2019)}]{Rudge2019}%
  \BibitemOpen
  \bibfield  {author} {\bibinfo {author} {\bibfnamefont {S.~L.}\ \bibnamefont
  {Rudge}}\ and\ \bibinfo {author} {\bibfnamefont {D.~S.}\ \bibnamefont
  {Kosov}},\ }\bibfield  {title} {\bibinfo {title} {Counting quantum jumps: A
  summary and comparison of fixed-time and fluctuating-time statistics in
  electron transport},\ }\href@noop {} {\bibfield  {journal} {\bibinfo
  {journal} {J. Chem. Phys.}\ }\textbf {\bibinfo {volume} {151}},\ \bibinfo
  {pages} {034107} (\bibinfo {year} {2019})}\BibitemShut {NoStop}%
\bibitem [{\citenamefont {Galperin}\ \emph {et~al.}(2006)\citenamefont
  {Galperin}, \citenamefont {Nitzan},\ and\ \citenamefont
  {Ratner}}]{GalperinShot2006}%
  \BibitemOpen
  \bibfield  {author} {\bibinfo {author} {\bibfnamefont {M.}~\bibnamefont
  {Galperin}}, \bibinfo {author} {\bibfnamefont {A.}~\bibnamefont {Nitzan}},\
  and\ \bibinfo {author} {\bibfnamefont {M.~A.}\ \bibnamefont {Ratner}},\
  }\bibfield  {title} {\bibinfo {title} {Inelastic tunneling effects on noise
  properties of molecular junctions},\ }\href@noop {} {\bibfield  {journal}
  {\bibinfo  {journal} {Phys. Rev. B}\ }\textbf {\bibinfo {volume} {74}},\
  \bibinfo {pages} {075326} (\bibinfo {year} {2006})}\BibitemShut {NoStop}%
\bibitem [{\citenamefont {Haug}\ and\ \citenamefont {Jauho}(2008)}]{Haug2008}%
  \BibitemOpen
  \bibfield  {author} {\bibinfo {author} {\bibfnamefont {H.}~\bibnamefont
  {Haug}}\ and\ \bibinfo {author} {\bibfnamefont {A.-P.}\ \bibnamefont
  {Jauho}},\ }\href {http://books.google.de/books?id=w1am24ZE9jQC} {\emph
  {\bibinfo {title} {Quantum Kinetics in Transport and Optics of
  Semiconductors}}},\ \bibinfo {series} {Springer Series in Solid-State
  Sciences}\ No.\ \bibinfo {number} {Bd. 6}\ (\bibinfo  {publisher}
  {Springer},\ \bibinfo {year} {2008})\BibitemShut {NoStop}%
\bibitem [{\citenamefont {Avriller}\ and\ \citenamefont
  {Levy~Yeyati}(2009)}]{Avriller2009}%
  \BibitemOpen
  \bibfield  {author} {\bibinfo {author} {\bibfnamefont {R.}~\bibnamefont
  {Avriller}}\ and\ \bibinfo {author} {\bibfnamefont {A.}~\bibnamefont
  {Levy~Yeyati}},\ }\bibfield  {title} {\bibinfo {title} {Electron-phonon
  interaction and full counting statistics in molecular junctions},\
  }\href@noop {} {\bibfield  {journal} {\bibinfo  {journal} {Phys. Rev. B}\
  }\textbf {\bibinfo {volume} {80}},\ \bibinfo {pages} {041309} (\bibinfo
  {year} {2009})}\BibitemShut {NoStop}%
\bibitem [{\citenamefont {Schmidt}\ and\ \citenamefont
  {Komnik}(2009)}]{Schmidt2009}%
  \BibitemOpen
  \bibfield  {author} {\bibinfo {author} {\bibfnamefont {T.~L.}\ \bibnamefont
  {Schmidt}}\ and\ \bibinfo {author} {\bibfnamefont {A.}~\bibnamefont
  {Komnik}},\ }\bibfield  {title} {\bibinfo {title} {Charge transfer statistics
  of a molecular quantum dot with a vibrational degree of freedom},\
  }\href@noop {} {\bibfield  {journal} {\bibinfo  {journal} {Phys. Rev. B}\
  }\textbf {\bibinfo {volume} {80}},\ \bibinfo {pages} {041307} (\bibinfo
  {year} {2009})}\BibitemShut {NoStop}%
\bibitem [{\citenamefont {Haupt}\ \emph {et~al.}(2010)\citenamefont {Haupt},
  \citenamefont {Novotn\'y},\ and\ \citenamefont {Belzig}}]{Haupt2010}%
  \BibitemOpen
  \bibfield  {author} {\bibinfo {author} {\bibfnamefont {F.}~\bibnamefont
  {Haupt}}, \bibinfo {author} {\bibfnamefont {T.~c.~v.}\ \bibnamefont
  {Novotn\'y}},\ and\ \bibinfo {author} {\bibfnamefont {W.}~\bibnamefont
  {Belzig}},\ }\bibfield  {title} {\bibinfo {title} {Current noise in molecular
  junctions: Effects of the electron-phonon interaction},\ }\href
  {https://doi.org/10.1103/PhysRevB.82.165441} {\bibfield  {journal} {\bibinfo
  {journal} {Phys. Rev. B}\ }\textbf {\bibinfo {volume} {82}},\ \bibinfo
  {pages} {165441} (\bibinfo {year} {2010})}\BibitemShut {NoStop}%
\bibitem [{\citenamefont {Novotn\'y}\ \emph {et~al.}(2011)\citenamefont
  {Novotn\'y}, \citenamefont {Haupt},\ and\ \citenamefont
  {Belzig}}]{Novotny2011}%
  \BibitemOpen
  \bibfield  {author} {\bibinfo {author} {\bibfnamefont {T.~c.~v.}\
  \bibnamefont {Novotn\'y}}, \bibinfo {author} {\bibfnamefont {F.}~\bibnamefont
  {Haupt}},\ and\ \bibinfo {author} {\bibfnamefont {W.}~\bibnamefont
  {Belzig}},\ }\bibfield  {title} {\bibinfo {title} {Nonequilibrium phonon
  backaction on the current noise in atomic-sized junctions},\ }\href
  {https://doi.org/10.1103/PhysRevB.84.113107} {\bibfield  {journal} {\bibinfo
  {journal} {Phys. Rev. B}\ }\textbf {\bibinfo {volume} {84}},\ \bibinfo
  {pages} {113107} (\bibinfo {year} {2011})}\BibitemShut {NoStop}%
\bibitem [{\citenamefont {Park}\ and\ \citenamefont
  {Galperin}(2011)}]{Park2011}%
  \BibitemOpen
  \bibfield  {author} {\bibinfo {author} {\bibfnamefont {T.-H.}\ \bibnamefont
  {Park}}\ and\ \bibinfo {author} {\bibfnamefont {M.}~\bibnamefont
  {Galperin}},\ }\bibfield  {title} {\bibinfo {title} {Self-consistent full
  counting statistics of inelastic transport},\ }\href
  {https://doi.org/10.1103/PhysRevB.84.205450} {\bibfield  {journal} {\bibinfo
  {journal} {Phys. Rev. B}\ }\textbf {\bibinfo {volume} {84}},\ \bibinfo
  {pages} {205450} (\bibinfo {year} {2011})}\BibitemShut {NoStop}%
\bibitem [{\citenamefont {Utsumi}\ \emph {et~al.}(2013)\citenamefont {Utsumi},
  \citenamefont {Entin-Wohlman}, \citenamefont {Ueda},\ and\ \citenamefont
  {Aharony}}]{Utsumi2013}%
  \BibitemOpen
  \bibfield  {author} {\bibinfo {author} {\bibfnamefont {Y.}~\bibnamefont
  {Utsumi}}, \bibinfo {author} {\bibfnamefont {O.}~\bibnamefont
  {Entin-Wohlman}}, \bibinfo {author} {\bibfnamefont {A.}~\bibnamefont
  {Ueda}},\ and\ \bibinfo {author} {\bibfnamefont {A.}~\bibnamefont
  {Aharony}},\ }\bibfield  {title} {\bibinfo {title} {Full-counting statistics
  for molecular junctions: Fluctuation theorem and singularities},\ }\href
  {https://doi.org/10.1103/PhysRevB.87.115407} {\bibfield  {journal} {\bibinfo
  {journal} {Phys. Rev. B}\ }\textbf {\bibinfo {volume} {87}},\ \bibinfo
  {pages} {115407} (\bibinfo {year} {2013})}\BibitemShut {NoStop}%
\bibitem [{\citenamefont {Seoane~Souto}\ \emph {et~al.}(2015)\citenamefont
  {Seoane~Souto}, \citenamefont {Avriller}, \citenamefont {Monreal},
  \citenamefont {Mart\'{\i}n-Rodero},\ and\ \citenamefont
  {Levy~Yeyati}}]{Souto2015}%
  \BibitemOpen
  \bibfield  {author} {\bibinfo {author} {\bibfnamefont {R.}~\bibnamefont
  {Seoane~Souto}}, \bibinfo {author} {\bibfnamefont {R.}~\bibnamefont
  {Avriller}}, \bibinfo {author} {\bibfnamefont {R.~C.}\ \bibnamefont
  {Monreal}}, \bibinfo {author} {\bibfnamefont {A.}~\bibnamefont
  {Mart\'{\i}n-Rodero}},\ and\ \bibinfo {author} {\bibfnamefont
  {A.}~\bibnamefont {Levy~Yeyati}},\ }\bibfield  {title} {\bibinfo {title}
  {Transient dynamics and waiting time distribution of molecular junctions in
  the polaronic regime},\ }\href {https://doi.org/10.1103/PhysRevB.92.125435}
  {\bibfield  {journal} {\bibinfo  {journal} {Phys. Rev. B}\ }\textbf {\bibinfo
  {volume} {92}},\ \bibinfo {pages} {125435} (\bibinfo {year}
  {2015})}\BibitemShut {NoStop}%
\bibitem [{\citenamefont {Miwa}\ \emph {et~al.}(2017)\citenamefont {Miwa},
  \citenamefont {Chen},\ and\ \citenamefont {Galperin}}]{Miwa2017}%
  \BibitemOpen
  \bibfield  {author} {\bibinfo {author} {\bibfnamefont {K.}~\bibnamefont
  {Miwa}}, \bibinfo {author} {\bibfnamefont {F.}~\bibnamefont {Chen}},\ and\
  \bibinfo {author} {\bibfnamefont {M.}~\bibnamefont {Galperin}},\ }\bibfield
  {title} {\bibinfo {title} {Towards noise simulation in interacting
  nonequilibrium systems strongly coupled to baths},\ }\href
  {https://www.scopus.com/inward/record.uri?eid=2-s2.0-85028454544&doi=10.1038%2fs41598-017-09060-0&partnerID=40&md5=2f748c6ea4efa0e2b5688e10a593f17a}
  {\bibfield  {journal} {\bibinfo  {journal} {Sci. Rep.}\ }\textbf {\bibinfo
  {volume} {7}} (\bibinfo {year} {2017})}\BibitemShut {NoStop}%
\bibitem [{\citenamefont {Dong}\ \emph {et~al.}(2017)\citenamefont {Dong},
  \citenamefont {Ding},\ and\ \citenamefont {Lei}}]{Dong2017}%
  \BibitemOpen
  \bibfield  {author} {\bibinfo {author} {\bibfnamefont {B.}~\bibnamefont
  {Dong}}, \bibinfo {author} {\bibfnamefont {G.~H.}\ \bibnamefont {Ding}},\
  and\ \bibinfo {author} {\bibfnamefont {X.~L.}\ \bibnamefont {Lei}},\
  }\bibfield  {title} {\bibinfo {title} {Full counting statistics of
  phonon-assisted andreev tunneling through a quantum dot coupled to normal and
  superconducting leads},\ }\href {https://doi.org/10.1103/PhysRevB.95.035409}
  {\bibfield  {journal} {\bibinfo  {journal} {Phys. Rev. B}\ }\textbf {\bibinfo
  {volume} {95}},\ \bibinfo {pages} {035409} (\bibinfo {year}
  {2017})}\BibitemShut {NoStop}%
\bibitem [{\citenamefont {Tang}\ \emph {et~al.}(2017)\citenamefont {Tang},
  \citenamefont {Xing},\ and\ \citenamefont {Wang}}]{Tang2017}%
  \BibitemOpen
  \bibfield  {author} {\bibinfo {author} {\bibfnamefont {G.}~\bibnamefont
  {Tang}}, \bibinfo {author} {\bibfnamefont {Y.}~\bibnamefont {Xing}},\ and\
  \bibinfo {author} {\bibfnamefont {J.}~\bibnamefont {Wang}},\ }\bibfield
  {title} {\bibinfo {title} {Short-time dynamics of molecular junctions after
  projective measurement},\ }\href {https://doi.org/10.1103/PhysRevB.96.075417}
  {\bibfield  {journal} {\bibinfo  {journal} {Phys. Rev. B}\ }\textbf {\bibinfo
  {volume} {96}},\ \bibinfo {pages} {075417} (\bibinfo {year}
  {2017})}\BibitemShut {NoStop}%
\bibitem [{\citenamefont {Stadler}\ \emph {et~al.}(2018)\citenamefont
  {Stadler}, \citenamefont {Rastelli},\ and\ \citenamefont
  {Belzig}}]{Stadler2018}%
  \BibitemOpen
  \bibfield  {author} {\bibinfo {author} {\bibfnamefont {P.}~\bibnamefont
  {Stadler}}, \bibinfo {author} {\bibfnamefont {G.}~\bibnamefont {Rastelli}},\
  and\ \bibinfo {author} {\bibfnamefont {W.}~\bibnamefont {Belzig}},\
  }\bibfield  {title} {\bibinfo {title} {Finite frequency current noise in the
  holstein model},\ }\href {https://doi.org/10.1103/PhysRevB.97.205408}
  {\bibfield  {journal} {\bibinfo  {journal} {Phys. Rev. B}\ }\textbf {\bibinfo
  {volume} {97}},\ \bibinfo {pages} {205408} (\bibinfo {year}
  {2018})}\BibitemShut {NoStop}%
\bibitem [{\citenamefont {Ridley}\ \emph {et~al.}(2018)\citenamefont {Ridley},
  \citenamefont {Singh}, \citenamefont {Gull},\ and\ \citenamefont
  {Cohen}}]{Ridley2018}%
  \BibitemOpen
  \bibfield  {author} {\bibinfo {author} {\bibfnamefont {M.}~\bibnamefont
  {Ridley}}, \bibinfo {author} {\bibfnamefont {V.~N.}\ \bibnamefont {Singh}},
  \bibinfo {author} {\bibfnamefont {E.}~\bibnamefont {Gull}},\ and\ \bibinfo
  {author} {\bibfnamefont {G.}~\bibnamefont {Cohen}},\ }\bibfield  {title}
  {\bibinfo {title} {Numerically exact full counting statistics of the
  nonequilibrium anderson impurity model},\ }\href
  {https://doi.org/10.1103/PhysRevB.97.115109} {\bibfield  {journal} {\bibinfo
  {journal} {Phys. Rev. B}\ }\textbf {\bibinfo {volume} {97}},\ \bibinfo
  {pages} {115109} (\bibinfo {year} {2018})}\BibitemShut {NoStop}%
\bibitem [{\citenamefont {H\"artle}\ \emph {et~al.}(2013)\citenamefont
  {H\"artle}, \citenamefont {Cohen}, \citenamefont {Reichman},\ and\
  \citenamefont {Millis}}]{Haertle2013a}%
  \BibitemOpen
  \bibfield  {author} {\bibinfo {author} {\bibfnamefont {R.}~\bibnamefont
  {H\"artle}}, \bibinfo {author} {\bibfnamefont {G.}~\bibnamefont {Cohen}},
  \bibinfo {author} {\bibfnamefont {D.~R.}\ \bibnamefont {Reichman}},\ and\
  \bibinfo {author} {\bibfnamefont {A.~J.}\ \bibnamefont {Millis}},\ }\bibfield
   {title} {\bibinfo {title} {Decoherence and lead-induced interdot coupling in
  nonequilibrium electron transport through interacting quantum dots: A
  hierarchical quantum master equation approach},\ }\href
  {https://doi.org/10.1103/PhysRevB.88.235426} {\bibfield  {journal} {\bibinfo
  {journal} {Phys. Rev. B}\ }\textbf {\bibinfo {volume} {88}},\ \bibinfo
  {pages} {235426} (\bibinfo {year} {2013})}\BibitemShut {NoStop}%
\bibitem [{\citenamefont {H\"artle}\ \emph {et~al.}(2015)\citenamefont
  {H\"artle}, \citenamefont {Cohen}, \citenamefont {Reichman},\ and\
  \citenamefont {Millis}}]{Haertle2015}%
  \BibitemOpen
  \bibfield  {author} {\bibinfo {author} {\bibfnamefont {R.}~\bibnamefont
  {H\"artle}}, \bibinfo {author} {\bibfnamefont {G.}~\bibnamefont {Cohen}},
  \bibinfo {author} {\bibfnamefont {D.~R.}\ \bibnamefont {Reichman}},\ and\
  \bibinfo {author} {\bibfnamefont {A.~J.}\ \bibnamefont {Millis}},\ }\bibfield
   {title} {\bibinfo {title} {Transport through an anderson impurity: Current
  ringing, nonlinear magnetization, and a direct comparison of continuous-time
  quantum monte carlo and hierarchical quantum master equations},\ }\href
  {https://doi.org/10.1103/PhysRevB.92.085430} {\bibfield  {journal} {\bibinfo
  {journal} {Phys. Rev. B}\ }\textbf {\bibinfo {volume} {92}},\ \bibinfo
  {pages} {085430} (\bibinfo {year} {2015})}\BibitemShut {NoStop}%
\bibitem [{\citenamefont {Tanimura}\ and\ \citenamefont
  {Kubo}(1989)}]{Tanimura1989}%
  \BibitemOpen
  \bibfield  {author} {\bibinfo {author} {\bibfnamefont {Y.}~\bibnamefont
  {Tanimura}}\ and\ \bibinfo {author} {\bibfnamefont {R.}~\bibnamefont
  {Kubo}},\ }\bibfield  {title} {\bibinfo {title} {Time evolution of a quantum
  system in contact with a nearly {Gauss}ian-markoffian noise bath},\ }\href
  {https://doi.org/10.1143/jpsj.58.101} {\bibfield  {journal} {\bibinfo
  {journal} {J. Phys. Soc. Jpn.}\ }\textbf {\bibinfo {volume} {58}},\ \bibinfo
  {pages} {101} (\bibinfo {year} {1989})}\BibitemShut {NoStop}%
\bibitem [{\citenamefont {Tanimura}(2006)}]{Tanimura2006}%
  \BibitemOpen
  \bibfield  {author} {\bibinfo {author} {\bibfnamefont {Y.}~\bibnamefont
  {Tanimura}},\ }\bibfield  {title} {\bibinfo {title} {Stochastic liouville,
  langevin, fokker-planck, and master equation approaches to quantum
  dissipative systems},\ }\href {https://doi.org/10.1143/jpsj.75.082001}
  {\bibfield  {journal} {\bibinfo  {journal} {J. Phys. Soc. Jpn.}\ }\textbf
  {\bibinfo {volume} {75}},\ \bibinfo {pages} {082001} (\bibinfo {year}
  {2006})}\BibitemShut {NoStop}%
\bibitem [{\citenamefont {Jin}\ \emph {et~al.}(2007)\citenamefont {Jin},
  \citenamefont {Welack}, \citenamefont {Luo}, \citenamefont {Li},
  \citenamefont {Cui}, \citenamefont {Xu},\ and\ \citenamefont
  {Yan}}]{Jin2007}%
  \BibitemOpen
  \bibfield  {author} {\bibinfo {author} {\bibfnamefont {J.}~\bibnamefont
  {Jin}}, \bibinfo {author} {\bibfnamefont {S.}~\bibnamefont {Welack}},
  \bibinfo {author} {\bibfnamefont {J.}~\bibnamefont {Luo}}, \bibinfo {author}
  {\bibfnamefont {X.-Q.}\ \bibnamefont {Li}}, \bibinfo {author} {\bibfnamefont
  {P.}~\bibnamefont {Cui}}, \bibinfo {author} {\bibfnamefont {R.-X.}\
  \bibnamefont {Xu}},\ and\ \bibinfo {author} {\bibfnamefont {Y.}~\bibnamefont
  {Yan}},\ }\bibfield  {title} {\bibinfo {title} {Dynamics of quantum
  dissipation systems interacting with fermion and boson grand canonical bath
  ensembles: Hierarchical equations of motion approach},\ }\href
  {https://doi.org/10.1063/1.2713104} {\bibfield  {journal} {\bibinfo
  {journal} {J. Chem. Phys.}\ }\textbf {\bibinfo {volume} {126}},\ \bibinfo
  {pages} {134113} (\bibinfo {year} {2007})}\BibitemShut {NoStop}%
\bibitem [{\citenamefont {Jin}\ \emph {et~al.}(2008)\citenamefont {Jin},
  \citenamefont {Zheng},\ and\ \citenamefont {Yan}}]{Jin2008}%
  \BibitemOpen
  \bibfield  {author} {\bibinfo {author} {\bibfnamefont {J.}~\bibnamefont
  {Jin}}, \bibinfo {author} {\bibfnamefont {X.}~\bibnamefont {Zheng}},\ and\
  \bibinfo {author} {\bibfnamefont {Y.}~\bibnamefont {Yan}},\ }\bibfield
  {title} {\bibinfo {title} {Exact dynamics of dissipative electronic systems
  and quantum transport: Hierarchical equations of motion approach},\ }\href
  {https://doi.org/10.1063/1.2938087} {\bibfield  {journal} {\bibinfo
  {journal} {J. Chem. Phys.}\ }\textbf {\bibinfo {volume} {128}},\ \bibinfo
  {pages} {234703} (\bibinfo {year} {2008})}\BibitemShut {NoStop}%
\bibitem [{\citenamefont {Zheng}\ \emph {et~al.}(2009)\citenamefont {Zheng},
  \citenamefont {Luo}, \citenamefont {Jin},\ and\ \citenamefont
  {Yan}}]{Zheng2009}%
  \BibitemOpen
  \bibfield  {author} {\bibinfo {author} {\bibfnamefont {X.}~\bibnamefont
  {Zheng}}, \bibinfo {author} {\bibfnamefont {J.}~\bibnamefont {Luo}}, \bibinfo
  {author} {\bibfnamefont {J.}~\bibnamefont {Jin}},\ and\ \bibinfo {author}
  {\bibfnamefont {Y.}~\bibnamefont {Yan}},\ }\bibfield  {title} {\bibinfo
  {title} {Complex non-{Markov}ian effect on time-dependent quantum
  transport},\ }\href {https://doi.org/10.1063/1.3095424} {\bibfield  {journal}
  {\bibinfo  {journal} {J. Chem. Phys.}\ }\textbf {\bibinfo {volume} {130}},\
  \bibinfo {pages} {124508} (\bibinfo {year} {2009})}\BibitemShut {NoStop}%
\bibitem [{\citenamefont {Li}\ \emph {et~al.}(2012)\citenamefont {Li},
  \citenamefont {Tong}, \citenamefont {Zheng}, \citenamefont {Hou},
  \citenamefont {Wei}, \citenamefont {Hu},\ and\ \citenamefont {Yan}}]{Li2012}%
  \BibitemOpen
  \bibfield  {author} {\bibinfo {author} {\bibfnamefont {Z.}~\bibnamefont
  {Li}}, \bibinfo {author} {\bibfnamefont {N.}~\bibnamefont {Tong}}, \bibinfo
  {author} {\bibfnamefont {X.}~\bibnamefont {Zheng}}, \bibinfo {author}
  {\bibfnamefont {D.}~\bibnamefont {Hou}}, \bibinfo {author} {\bibfnamefont
  {J.}~\bibnamefont {Wei}}, \bibinfo {author} {\bibfnamefont {J.}~\bibnamefont
  {Hu}},\ and\ \bibinfo {author} {\bibfnamefont {Y.}~\bibnamefont {Yan}},\
  }\bibfield  {title} {\bibinfo {title} {Hierarchical liouville-space approach
  for accurate and universal characterization of quantum impurity systems},\
  }\href {https://doi.org/10.1103/PhysRevLett.109.266403} {\bibfield  {journal}
  {\bibinfo  {journal} {Phys. Rev. Lett.}\ }\textbf {\bibinfo {volume} {109}},\
  \bibinfo {pages} {266403} (\bibinfo {year} {2012})}\BibitemShut {NoStop}%
\bibitem [{\citenamefont {Zheng}\ \emph {et~al.}(2013)\citenamefont {Zheng},
  \citenamefont {Yan},\ and\ \citenamefont {Di~Ventra}}]{Zheng2013}%
  \BibitemOpen
  \bibfield  {author} {\bibinfo {author} {\bibfnamefont {X.}~\bibnamefont
  {Zheng}}, \bibinfo {author} {\bibfnamefont {Y.}~\bibnamefont {Yan}},\ and\
  \bibinfo {author} {\bibfnamefont {M.}~\bibnamefont {Di~Ventra}},\ }\bibfield
  {title} {\bibinfo {title} {Kondo memory in driven strongly correlated quantum
  dots},\ }\href {https://doi.org/10.1103/PhysRevLett.111.086601} {\bibfield
  {journal} {\bibinfo  {journal} {Phys. Rev. Lett.}\ }\textbf {\bibinfo
  {volume} {111}},\ \bibinfo {pages} {086601} (\bibinfo {year}
  {2013})}\BibitemShut {NoStop}%
\bibitem [{\citenamefont {Cheng}\ \emph {et~al.}(2015)\citenamefont {Cheng},
  \citenamefont {Wei},\ and\ \citenamefont {Yan}}]{Cheng2015}%
  \BibitemOpen
  \bibfield  {author} {\bibinfo {author} {\bibfnamefont {Y.}~\bibnamefont
  {Cheng}}, \bibinfo {author} {\bibfnamefont {J.}~\bibnamefont {Wei}},\ and\
  \bibinfo {author} {\bibfnamefont {Y.}~\bibnamefont {Yan}},\ }\bibfield
  {title} {\bibinfo {title} {Reappearance of the kondo effect in serially
  coupled symmetric triple quantum dots},\ }\href
  {https://doi.org/10.1209/0295-5075/112/57001} {\bibfield  {journal} {\bibinfo
   {journal} {Europhys. Lett.}\ }\textbf {\bibinfo {volume} {112}},\ \bibinfo
  {pages} {57001} (\bibinfo {year} {2015})}\BibitemShut {NoStop}%
\bibitem [{\citenamefont {Ye}\ \emph {et~al.}(2016)\citenamefont {Ye},
  \citenamefont {Wang}, \citenamefont {Hou}, \citenamefont {Xu}, \citenamefont
  {Zheng},\ and\ \citenamefont {Yan}}]{Ye2016}%
  \BibitemOpen
  \bibfield  {author} {\bibinfo {author} {\bibfnamefont {L.}~\bibnamefont
  {Ye}}, \bibinfo {author} {\bibfnamefont {X.}~\bibnamefont {Wang}}, \bibinfo
  {author} {\bibfnamefont {D.}~\bibnamefont {Hou}}, \bibinfo {author}
  {\bibfnamefont {R.-X.}\ \bibnamefont {Xu}}, \bibinfo {author} {\bibfnamefont
  {X.}~\bibnamefont {Zheng}},\ and\ \bibinfo {author} {\bibfnamefont
  {Y.}~\bibnamefont {Yan}},\ }\bibfield  {title} {\bibinfo {title} {Heom-quick:
  a program for accurate, efficient, and universal characterization of strongly
  correlated quantum impurity systems},\ }\href
  {https://doi.org/10.1002/wcms.1269} {\bibfield  {journal} {\bibinfo
  {journal} {WIREs Comput. Mol. Sci.}\ }\textbf {\bibinfo {volume} {6}},\
  \bibinfo {pages} {608} (\bibinfo {year} {2016})}\BibitemShut {NoStop}%
\bibitem [{\citenamefont {Cheng}\ \emph {et~al.}(2017)\citenamefont {Cheng},
  \citenamefont {Wang}, \citenamefont {Wei}, \citenamefont {Zhu},\ and\
  \citenamefont {Yan}}]{Cheng2017}%
  \BibitemOpen
  \bibfield  {author} {\bibinfo {author} {\bibfnamefont {Y.}~\bibnamefont
  {Cheng}}, \bibinfo {author} {\bibfnamefont {Y.}~\bibnamefont {Wang}},
  \bibinfo {author} {\bibfnamefont {J.}~\bibnamefont {Wei}}, \bibinfo {author}
  {\bibfnamefont {Z.}~\bibnamefont {Zhu}},\ and\ \bibinfo {author}
  {\bibfnamefont {Y.}~\bibnamefont {Yan}},\ }\bibfield  {title} {\bibinfo
  {title} {Long-range exchange interaction in triple quantum dots in the kondo
  regime},\ }\href {https://doi.org/10.1103/PhysRevB.95.155417} {\bibfield
  {journal} {\bibinfo  {journal} {Phys. Rev. B}\ }\textbf {\bibinfo {volume}
  {95}},\ \bibinfo {pages} {155417} (\bibinfo {year} {2017})}\BibitemShut
  {NoStop}%
\bibitem [{\citenamefont {Li}\ \emph {et~al.}(2017)\citenamefont {Li},
  \citenamefont {Wei}, \citenamefont {Zheng}, \citenamefont {Yan},\ and\
  \citenamefont {Luo}}]{Li2017}%
  \BibitemOpen
  \bibfield  {author} {\bibinfo {author} {\bibfnamefont {Z.}~\bibnamefont
  {Li}}, \bibinfo {author} {\bibfnamefont {J.}~\bibnamefont {Wei}}, \bibinfo
  {author} {\bibfnamefont {X.}~\bibnamefont {Zheng}}, \bibinfo {author}
  {\bibfnamefont {Y.}~\bibnamefont {Yan}},\ and\ \bibinfo {author}
  {\bibfnamefont {H.-G.}\ \bibnamefont {Luo}},\ }\bibfield  {title} {\bibinfo
  {title} {Corrected kondo temperature beyond the conventional kondo scaling
  limit},\ }\href {http://stacks.iop.org/0953-8984/29/i=17/a=175601} {\bibfield
   {journal} {\bibinfo  {journal} {J. Phys.: Condens. Matter}\ }\textbf
  {\bibinfo {volume} {29}},\ \bibinfo {pages} {175601} (\bibinfo {year}
  {2017})}\BibitemShut {NoStop}%
\bibitem [{\citenamefont {Hou}\ \emph {et~al.}(2017{\natexlab{a}})\citenamefont
  {Hou}, \citenamefont {Wang}, \citenamefont {Wei}, \citenamefont {Zhu},\ and\
  \citenamefont {Yan}}]{Hou2017}%
  \BibitemOpen
  \bibfield  {author} {\bibinfo {author} {\bibfnamefont {W.}~\bibnamefont
  {Hou}}, \bibinfo {author} {\bibfnamefont {Y.}~\bibnamefont {Wang}}, \bibinfo
  {author} {\bibfnamefont {J.}~\bibnamefont {Wei}}, \bibinfo {author}
  {\bibfnamefont {Z.}~\bibnamefont {Zhu}},\ and\ \bibinfo {author}
  {\bibfnamefont {Y.}~\bibnamefont {Yan}},\ }\bibfield  {title} {\bibinfo
  {title} {Many-body tunneling and nonequilibrium dynamics of doublons in
  strongly correlated quantum dots},\ }\href
  {https://doi.org/10.1038/s41598-017-02728-7} {\bibfield  {journal} {\bibinfo
  {journal} {Sci. Rep.}\ }\textbf {\bibinfo {volume} {7}},\ \bibinfo {pages}
  {2486} (\bibinfo {year} {2017}{\natexlab{a}})}\BibitemShut {NoStop}%
\bibitem [{\citenamefont {Hou}\ \emph {et~al.}(2017{\natexlab{b}})\citenamefont
  {Hou}, \citenamefont {Wang}, \citenamefont {Wei},\ and\ \citenamefont
  {Yan}}]{Hou2017a}%
  \BibitemOpen
  \bibfield  {author} {\bibinfo {author} {\bibfnamefont {W.}~\bibnamefont
  {Hou}}, \bibinfo {author} {\bibfnamefont {Y.}~\bibnamefont {Wang}}, \bibinfo
  {author} {\bibfnamefont {J.}~\bibnamefont {Wei}},\ and\ \bibinfo {author}
  {\bibfnamefont {Y.}~\bibnamefont {Yan}},\ }\bibfield  {title} {\bibinfo
  {title} {Manipulation of pauli spin blockade in double quantum dot systems},\
  }\href {https://doi.org/10.1063/1.4985146} {\bibfield  {journal} {\bibinfo
  {journal} {J. Chem. Phys.}\ }\textbf {\bibinfo {volume} {146}},\ \bibinfo
  {pages} {224304} (\bibinfo {year} {2017}{\natexlab{b}})}\BibitemShut
  {NoStop}%
\bibitem [{\citenamefont {H\"artle}\ and\ \citenamefont
  {Millis}(2014)}]{Haertle2014}%
  \BibitemOpen
  \bibfield  {author} {\bibinfo {author} {\bibfnamefont {R.}~\bibnamefont
  {H\"artle}}\ and\ \bibinfo {author} {\bibfnamefont {A.~J.}\ \bibnamefont
  {Millis}},\ }\bibfield  {title} {\bibinfo {title} {Formation of
  nonequilibrium steady states in interacting double quantum dots: When
  coherences dominate the charge distribution},\ }\href
  {https://doi.org/10.1103/PhysRevB.90.245426} {\bibfield  {journal} {\bibinfo
  {journal} {Phys. Rev. B}\ }\textbf {\bibinfo {volume} {90}},\ \bibinfo
  {pages} {245426} (\bibinfo {year} {2014})}\BibitemShut {NoStop}%
\bibitem [{\citenamefont {Wenderoth}\ \emph {et~al.}(2016)\citenamefont
  {Wenderoth}, \citenamefont {B\"atge},\ and\ \citenamefont
  {H\"artle}}]{Wenderoth2016}%
  \BibitemOpen
  \bibfield  {author} {\bibinfo {author} {\bibfnamefont {S.}~\bibnamefont
  {Wenderoth}}, \bibinfo {author} {\bibfnamefont {J.}~\bibnamefont {B\"atge}},\
  and\ \bibinfo {author} {\bibfnamefont {R.}~\bibnamefont {H\"artle}},\
  }\bibfield  {title} {\bibinfo {title} {Sharp peaks in the conductance of a
  double quantum dot and a quantum-dot spin valve at high temperatures: A
  hierarchical quantum master equation approach},\ }\href
  {https://doi.org/10.1103/PhysRevB.94.121303} {\bibfield  {journal} {\bibinfo
  {journal} {Phys. Rev. B}\ }\textbf {\bibinfo {volume} {94}},\ \bibinfo
  {pages} {121303} (\bibinfo {year} {2016})}\BibitemShut {NoStop}%
\bibitem [{\citenamefont {Schinabeck}\ \emph {et~al.}(2016)\citenamefont
  {Schinabeck}, \citenamefont {Erpenbeck}, \citenamefont {H\"artle},\ and\
  \citenamefont {Thoss}}]{Schinabeck2016}%
  \BibitemOpen
  \bibfield  {author} {\bibinfo {author} {\bibfnamefont {C.}~\bibnamefont
  {Schinabeck}}, \bibinfo {author} {\bibfnamefont {A.}~\bibnamefont
  {Erpenbeck}}, \bibinfo {author} {\bibfnamefont {R.}~\bibnamefont
  {H\"artle}},\ and\ \bibinfo {author} {\bibfnamefont {M.}~\bibnamefont
  {Thoss}},\ }\bibfield  {title} {\bibinfo {title} {Hierarchical quantum master
  equation approach to electronic-vibrational coupling in nonequilibrium
  transport through nanosystems},\ }\href
  {https://doi.org/10.1103/PhysRevB.94.201407} {\bibfield  {journal} {\bibinfo
  {journal} {Phys. Rev. B}\ }\textbf {\bibinfo {volume} {94}},\ \bibinfo
  {pages} {201407} (\bibinfo {year} {2016})}\BibitemShut {NoStop}%
\bibitem [{\citenamefont {Schinabeck}\ \emph {et~al.}(2018)\citenamefont
  {Schinabeck}, \citenamefont {H\"artle},\ and\ \citenamefont
  {Thoss}}]{Schinabeck2018}%
  \BibitemOpen
  \bibfield  {author} {\bibinfo {author} {\bibfnamefont {C.}~\bibnamefont
  {Schinabeck}}, \bibinfo {author} {\bibfnamefont {R.}~\bibnamefont
  {H\"artle}},\ and\ \bibinfo {author} {\bibfnamefont {M.}~\bibnamefont
  {Thoss}},\ }\bibfield  {title} {\bibinfo {title} {Hierarchical quantum master
  equation approach to electronic-vibrational coupling in nonequilibrium
  transport through nanosystems: Reservoir formulation and application to
  vibrational instabilities},\ }\href
  {https://doi.org/10.1103/PhysRevB.97.235429} {\bibfield  {journal} {\bibinfo
  {journal} {Phys. Rev. B}\ }\textbf {\bibinfo {volume} {97}},\ \bibinfo
  {pages} {235429} (\bibinfo {year} {2018})}\BibitemShut {NoStop}%
\bibitem [{\citenamefont {Flindt}\ \emph {et~al.}(2010)\citenamefont {Flindt},
  \citenamefont {Novotn\'y}, \citenamefont {Braggio},\ and\ \citenamefont
  {Jauho}}]{Flindt2010}%
  \BibitemOpen
  \bibfield  {author} {\bibinfo {author} {\bibfnamefont {C.}~\bibnamefont
  {Flindt}}, \bibinfo {author} {\bibfnamefont {T.}~\bibnamefont {Novotn\'y}},
  \bibinfo {author} {\bibfnamefont {A.}~\bibnamefont {Braggio}},\ and\ \bibinfo
  {author} {\bibfnamefont {A.-P.}\ \bibnamefont {Jauho}},\ }\bibfield  {title}
  {\bibinfo {title} {Counting statistics of transport through coulomb blockade
  nanostructures: High-order cumulants and non-markovian effects},\ }\href
  {https://doi.org/10.1103/PhysRevB.82.155407} {\bibfield  {journal} {\bibinfo
  {journal} {Phys. Rev. B}\ }\textbf {\bibinfo {volume} {82}},\ \bibinfo
  {pages} {155407} (\bibinfo {year} {2010})}\BibitemShut {NoStop}%
\bibitem [{\citenamefont {Wang}\ \emph {et~al.}(2013)\citenamefont {Wang},
  \citenamefont {Zheng}, \citenamefont {Jin},\ and\ \citenamefont
  {Yan}}]{Wang2013}%
  \BibitemOpen
  \bibfield  {author} {\bibinfo {author} {\bibfnamefont {S.}~\bibnamefont
  {Wang}}, \bibinfo {author} {\bibfnamefont {X.}~\bibnamefont {Zheng}},
  \bibinfo {author} {\bibfnamefont {J.}~\bibnamefont {Jin}},\ and\ \bibinfo
  {author} {\bibfnamefont {Y.}~\bibnamefont {Yan}},\ }\bibfield  {title}
  {\bibinfo {title} {Hierarchical liouville-space approach to nonequilibrium
  dynamical properties of quantum impurity systems},\ }\href
  {https://doi.org/10.1103/PhysRevB.88.035129} {\bibfield  {journal} {\bibinfo
  {journal} {Phys. Rev. B}\ }\textbf {\bibinfo {volume} {88}},\ \bibinfo
  {pages} {035129} (\bibinfo {year} {2013})}\BibitemShut {NoStop}%
\bibitem [{\citenamefont {Secker}\ \emph {et~al.}(2011)\citenamefont {Secker},
  \citenamefont {Wagner}, \citenamefont {Ballmann}, \citenamefont {H\"artle},
  \citenamefont {Thoss},\ and\ \citenamefont {Weber}}]{Secker2011}%
  \BibitemOpen
  \bibfield  {author} {\bibinfo {author} {\bibfnamefont {D.}~\bibnamefont
  {Secker}}, \bibinfo {author} {\bibfnamefont {S.}~\bibnamefont {Wagner}},
  \bibinfo {author} {\bibfnamefont {S.}~\bibnamefont {Ballmann}}, \bibinfo
  {author} {\bibfnamefont {R.}~\bibnamefont {H\"artle}}, \bibinfo {author}
  {\bibfnamefont {M.}~\bibnamefont {Thoss}},\ and\ \bibinfo {author}
  {\bibfnamefont {H.~B.}\ \bibnamefont {Weber}},\ }\bibfield  {title} {\bibinfo
  {title} {Resonant vibrations, peak broadening, and noise in single molecule
  contacts: The nature of the first conductance peak},\ }\href
  {https://doi.org/10.1103/PhysRevLett.106.136807} {\bibfield  {journal}
  {\bibinfo  {journal} {Phys. Rev. Lett.}\ }\textbf {\bibinfo {volume} {106}},\
  \bibinfo {pages} {136807} (\bibinfo {year} {2011})}\BibitemShut {NoStop}%
\bibitem [{\citenamefont {N\'eel}\ \emph {et~al.}(2011)\citenamefont {N\'eel},
  \citenamefont {Kr\"oger},\ and\ \citenamefont {Berndt}}]{Neel2011}%
  \BibitemOpen
  \bibfield  {author} {\bibinfo {author} {\bibfnamefont {N.}~\bibnamefont
  {N\'eel}}, \bibinfo {author} {\bibfnamefont {J.}~\bibnamefont {Kr\"oger}},\
  and\ \bibinfo {author} {\bibfnamefont {R.}~\bibnamefont {Berndt}},\
  }\bibfield  {title} {\bibinfo {title} {Two-level conductance fluctuations of
  a single-molecule junction},\ }\href {https://doi.org/10.1021/nl201327c}
  {\bibfield  {journal} {\bibinfo  {journal} {Nano Lett.}\ }\textbf {\bibinfo
  {volume} {11}},\ \bibinfo {pages} {3593} (\bibinfo {year}
  {2011})}\BibitemShut {NoStop}%
\bibitem [{\citenamefont {Lau}\ \emph {et~al.}(2016)\citenamefont {Lau},
  \citenamefont {Sadeghi}, \citenamefont {Rogers}, \citenamefont {Sangtarash},
  \citenamefont {Dallas}, \citenamefont {Porfyrakis}, \citenamefont {Warner},
  \citenamefont {Lambert}, \citenamefont {Briggs},\ and\ \citenamefont
  {Mol}}]{Lau2016}%
  \BibitemOpen
  \bibfield  {author} {\bibinfo {author} {\bibfnamefont {C.~S.}\ \bibnamefont
  {Lau}}, \bibinfo {author} {\bibfnamefont {H.}~\bibnamefont {Sadeghi}},
  \bibinfo {author} {\bibfnamefont {G.}~\bibnamefont {Rogers}}, \bibinfo
  {author} {\bibfnamefont {S.}~\bibnamefont {Sangtarash}}, \bibinfo {author}
  {\bibfnamefont {P.}~\bibnamefont {Dallas}}, \bibinfo {author} {\bibfnamefont
  {K.}~\bibnamefont {Porfyrakis}}, \bibinfo {author} {\bibfnamefont
  {J.}~\bibnamefont {Warner}}, \bibinfo {author} {\bibfnamefont {C.~J.}\
  \bibnamefont {Lambert}}, \bibinfo {author} {\bibfnamefont {G.~A.~D.}\
  \bibnamefont {Briggs}},\ and\ \bibinfo {author} {\bibfnamefont {J.~A.}\
  \bibnamefont {Mol}},\ }\bibfield  {title} {\bibinfo {title} {Redox-dependent
  franck-condon blockade and avalanche transport in a graphene-fullerene
  single-molecule transistor},\ }\href
  {https://doi.org/10.1021/acs.nanolett.5b03434} {\bibfield  {journal}
  {\bibinfo  {journal} {Nano Lett.}\ }\textbf {\bibinfo {volume} {16}},\
  \bibinfo {pages} {170} (\bibinfo {year} {2016})}\BibitemShut {NoStop}%
\bibitem [{\citenamefont {Weig}\ \emph {et~al.}(2004)\citenamefont {Weig},
  \citenamefont {Blick}, \citenamefont {Brandes}, \citenamefont {Kirschbaum},
  \citenamefont {Wegscheider}, \citenamefont {Bichler},\ and\ \citenamefont
  {Kotthaus}}]{Weig2004}%
  \BibitemOpen
  \bibfield  {author} {\bibinfo {author} {\bibfnamefont {E.~M.}\ \bibnamefont
  {Weig}}, \bibinfo {author} {\bibfnamefont {R.~H.}\ \bibnamefont {Blick}},
  \bibinfo {author} {\bibfnamefont {T.}~\bibnamefont {Brandes}}, \bibinfo
  {author} {\bibfnamefont {J.}~\bibnamefont {Kirschbaum}}, \bibinfo {author}
  {\bibfnamefont {W.}~\bibnamefont {Wegscheider}}, \bibinfo {author}
  {\bibfnamefont {M.}~\bibnamefont {Bichler}},\ and\ \bibinfo {author}
  {\bibfnamefont {J.~P.}\ \bibnamefont {Kotthaus}},\ }\bibfield  {title}
  {\bibinfo {title} {Single-electron-phonon interaction in a suspended quantum
  dot phonon cavity},\ }\href {https://doi.org/10.1103/PhysRevLett.92.046804}
  {\bibfield  {journal} {\bibinfo  {journal} {Phys. Rev. Lett.}\ }\textbf
  {\bibinfo {volume} {92}},\ \bibinfo {pages} {046804} (\bibinfo {year}
  {2004})}\BibitemShut {NoStop}%
\bibitem [{\citenamefont {Sapmaz}\ \emph {et~al.}(2006)\citenamefont {Sapmaz},
  \citenamefont {Jarillo-Herrero}, \citenamefont {Blanter}, \citenamefont
  {Dekker},\ and\ \citenamefont {van~der Zant}}]{Sapmaz2006}%
  \BibitemOpen
  \bibfield  {author} {\bibinfo {author} {\bibfnamefont {S.}~\bibnamefont
  {Sapmaz}}, \bibinfo {author} {\bibfnamefont {P.}~\bibnamefont
  {Jarillo-Herrero}}, \bibinfo {author} {\bibfnamefont {Y.~M.}\ \bibnamefont
  {Blanter}}, \bibinfo {author} {\bibfnamefont {C.}~\bibnamefont {Dekker}},\
  and\ \bibinfo {author} {\bibfnamefont {H.~S.~J.}\ \bibnamefont {van~der
  Zant}},\ }\bibfield  {title} {\bibinfo {title} {Tunneling in suspended carbon
  nanotubes assisted by longitudinal phonons},\ }\href
  {https://doi.org/10.1103/PhysRevLett.96.026801} {\bibfield  {journal}
  {\bibinfo  {journal} {Phys. Rev. Lett.}\ }\textbf {\bibinfo {volume} {96}},\
  \bibinfo {pages} {026801} (\bibinfo {year} {2006})}\BibitemShut {NoStop}%
\bibitem [{\citenamefont {Leturcq}\ \emph {et~al.}(2009)\citenamefont
  {Leturcq}, \citenamefont {Stampfer}, \citenamefont {Inderbitzin},
  \citenamefont {Durrer}, \citenamefont {Hierold}, \citenamefont {Mariani},
  \citenamefont {Schultz}, \citenamefont {von Oppen},\ and\ \citenamefont
  {Ensslin}}]{Leturcq2009}%
  \BibitemOpen
  \bibfield  {author} {\bibinfo {author} {\bibfnamefont {R.}~\bibnamefont
  {Leturcq}}, \bibinfo {author} {\bibfnamefont {C.}~\bibnamefont {Stampfer}},
  \bibinfo {author} {\bibfnamefont {K.}~\bibnamefont {Inderbitzin}}, \bibinfo
  {author} {\bibfnamefont {L.}~\bibnamefont {Durrer}}, \bibinfo {author}
  {\bibfnamefont {C.}~\bibnamefont {Hierold}}, \bibinfo {author} {\bibfnamefont
  {E.}~\bibnamefont {Mariani}}, \bibinfo {author} {\bibfnamefont {M.~G.}\
  \bibnamefont {Schultz}}, \bibinfo {author} {\bibfnamefont {F.}~\bibnamefont
  {von Oppen}},\ and\ \bibinfo {author} {\bibfnamefont {K.}~\bibnamefont
  {Ensslin}},\ }\bibfield  {title} {\bibinfo {title} {Franck{\textendash}condon
  blockade in suspended carbon nanotube quantum dots},\ }\href
  {https://doi.org/10.1038/nphys1234} {\bibfield  {journal} {\bibinfo
  {journal} {Nat. Phys.}\ }\textbf {\bibinfo {volume} {5}},\ \bibinfo {pages}
  {327} (\bibinfo {year} {2009})}\BibitemShut {NoStop}%
\bibitem [{\citenamefont {van~der Molen}\ and\ \citenamefont
  {Liljeroth}(2010)}]{Molen2010}%
  \BibitemOpen
  \bibfield  {author} {\bibinfo {author} {\bibfnamefont {S.~J.}\ \bibnamefont
  {van~der Molen}}\ and\ \bibinfo {author} {\bibfnamefont {P.}~\bibnamefont
  {Liljeroth}},\ }\bibfield  {title} {\bibinfo {title} {Charge transport
  through molecular switches},\ }\href
  {http://stacks.iop.org/0953-8984/22/i=13/a=133001} {\bibfield  {journal}
  {\bibinfo  {journal} {J. Phys.: Condens. Matter}\ }\textbf {\bibinfo {volume}
  {22}},\ \bibinfo {pages} {133001} (\bibinfo {year} {2010})}\BibitemShut
  {NoStop}%
\bibitem [{\citenamefont {Gaudioso}\ \emph {et~al.}(2000)\citenamefont
  {Gaudioso}, \citenamefont {Lauhon},\ and\ \citenamefont {Ho}}]{Gaudioso2000}%
  \BibitemOpen
  \bibfield  {author} {\bibinfo {author} {\bibfnamefont {J.}~\bibnamefont
  {Gaudioso}}, \bibinfo {author} {\bibfnamefont {L.~J.}\ \bibnamefont
  {Lauhon}},\ and\ \bibinfo {author} {\bibfnamefont {W.}~\bibnamefont {Ho}},\
  }\bibfield  {title} {\bibinfo {title} {Vibrationally mediated negative
  differential resistance in a single molecule},\ }\href@noop {} {\bibfield
  {journal} {\bibinfo  {journal} {Phys. Rev. Lett.}\ }\textbf {\bibinfo
  {volume} {85}},\ \bibinfo {pages} {1918} (\bibinfo {year}
  {2000})}\BibitemShut {NoStop}%
\bibitem [{\citenamefont {Pop}\ \emph {et~al.}(2005)\citenamefont {Pop},
  \citenamefont {Mann}, \citenamefont {Cao}, \citenamefont {Wang},
  \citenamefont {Goodson},\ and\ \citenamefont {Dai}}]{Pop2005}%
  \BibitemOpen
  \bibfield  {author} {\bibinfo {author} {\bibfnamefont {E.}~\bibnamefont
  {Pop}}, \bibinfo {author} {\bibfnamefont {D.}~\bibnamefont {Mann}}, \bibinfo
  {author} {\bibfnamefont {J.}~\bibnamefont {Cao}}, \bibinfo {author}
  {\bibfnamefont {Q.}~\bibnamefont {Wang}}, \bibinfo {author} {\bibfnamefont
  {K.}~\bibnamefont {Goodson}},\ and\ \bibinfo {author} {\bibfnamefont
  {H.}~\bibnamefont {Dai}},\ }\bibfield  {title} {\bibinfo {title} {Negative
  differential conductance and hot phonons in suspended nanotube molecular
  wires},\ }\href {https://doi.org/10.1103/PhysRevLett.95.155505} {\bibfield
  {journal} {\bibinfo  {journal} {Phys. Rev. Lett.}\ }\textbf {\bibinfo
  {volume} {95}},\ \bibinfo {pages} {155505} (\bibinfo {year}
  {2005})}\BibitemShut {NoStop}%
\bibitem [{\citenamefont {Galperin}\ \emph {et~al.}(2005)\citenamefont
  {Galperin}, \citenamefont {Ratner},\ and\ \citenamefont
  {Nitzan}}]{Galperin2005}%
  \BibitemOpen
  \bibfield  {author} {\bibinfo {author} {\bibfnamefont {M.}~\bibnamefont
  {Galperin}}, \bibinfo {author} {\bibfnamefont {M.~A.}\ \bibnamefont
  {Ratner}},\ and\ \bibinfo {author} {\bibfnamefont {A.}~\bibnamefont
  {Nitzan}},\ }\bibfield  {title} {\bibinfo {title} {Hysteresis, switching, and
  negative differential resistance in molecular junctions: A polaron model},\
  }\href {https://doi.org/10.1021/nl048216c} {\bibfield  {journal} {\bibinfo
  {journal} {Nano Lett.}\ }\textbf {\bibinfo {volume} {5}},\ \bibinfo {pages}
  {125} (\bibinfo {year} {2005})}\BibitemShut {NoStop}%
\bibitem [{\citenamefont {Leijnse}\ and\ \citenamefont
  {Wegewijs}(2008)}]{Leijnse2008}%
  \BibitemOpen
  \bibfield  {author} {\bibinfo {author} {\bibfnamefont {M.}~\bibnamefont
  {Leijnse}}\ and\ \bibinfo {author} {\bibfnamefont {M.~R.}\ \bibnamefont
  {Wegewijs}},\ }\bibfield  {title} {\bibinfo {title} {Kinetic equations for
  transport through single-molecule transistors},\ }\href
  {https://doi.org/10.1103/PhysRevB.78.235424} {\bibfield  {journal} {\bibinfo
  {journal} {Phys. Rev. B}\ }\textbf {\bibinfo {volume} {78}},\ \bibinfo
  {pages} {235424} (\bibinfo {year} {2008})}\BibitemShut {NoStop}%
\bibitem [{\citenamefont {H\"artle}\ and\ \citenamefont
  {Thoss}(2011{\natexlab{a}})}]{Haertle2011}%
  \BibitemOpen
  \bibfield  {author} {\bibinfo {author} {\bibfnamefont {R.}~\bibnamefont
  {H\"artle}}\ and\ \bibinfo {author} {\bibfnamefont {M.}~\bibnamefont
  {Thoss}},\ }\bibfield  {title} {\bibinfo {title} {Resonant electron transport
  in single-molecule junctions: Vibrational excitation, rectification, negative
  differential resistance, and local cooling},\ }\href
  {https://doi.org/10.1103/PhysRevB.83.115414} {\bibfield  {journal} {\bibinfo
  {journal} {Phys. Rev. B}\ }\textbf {\bibinfo {volume} {83}},\ \bibinfo
  {pages} {115414} (\bibinfo {year} {2011}{\natexlab{a}})}\BibitemShut
  {NoStop}%
\bibitem [{\citenamefont {Koch}\ and\ \citenamefont {von
  Oppen}(2005)}]{Koch2005}%
  \BibitemOpen
  \bibfield  {author} {\bibinfo {author} {\bibfnamefont {J.}~\bibnamefont
  {Koch}}\ and\ \bibinfo {author} {\bibfnamefont {F.}~\bibnamefont {von
  Oppen}},\ }\bibfield  {title} {\bibinfo {title} {Franck-condon blockade and
  giant fano factors in transport through single molecules},\ }\href
  {https://doi.org/10.1103/PhysRevLett.94.206804} {\bibfield  {journal}
  {\bibinfo  {journal} {Phys. Rev. Lett.}\ }\textbf {\bibinfo {volume} {94}},\
  \bibinfo {pages} {206804} (\bibinfo {year} {2005})}\BibitemShut {NoStop}%
\bibitem [{\citenamefont {H\"artle}\ \emph {et~al.}(2011)\citenamefont
  {H\"artle}, \citenamefont {Butzin}, \citenamefont {Rubio-Pons},\ and\
  \citenamefont {Thoss}}]{Haertle2011b}%
  \BibitemOpen
  \bibfield  {author} {\bibinfo {author} {\bibfnamefont {R.}~\bibnamefont
  {H\"artle}}, \bibinfo {author} {\bibfnamefont {M.}~\bibnamefont {Butzin}},
  \bibinfo {author} {\bibfnamefont {O.}~\bibnamefont {Rubio-Pons}},\ and\
  \bibinfo {author} {\bibfnamefont {M.}~\bibnamefont {Thoss}},\ }\bibfield
  {title} {\bibinfo {title} {Quantum interference and decoherence in
  single-molecule junctions: How vibrations induce electrical current},\
  }\href@noop {} {\bibfield  {journal} {\bibinfo  {journal} {Phys. Rev. Lett.}\
  }\textbf {\bibinfo {volume} {107}},\ \bibinfo {pages} {046802} (\bibinfo
  {year} {2011})}\BibitemShut {NoStop}%
\bibitem [{\citenamefont {Ballmann}\ \emph {et~al.}(2012)\citenamefont
  {Ballmann}, \citenamefont {H\"artle}, \citenamefont {Coto}, \citenamefont
  {Elbing}, \citenamefont {Mayor}, \citenamefont {Bryce}, \citenamefont
  {Thoss},\ and\ \citenamefont {Weber}}]{Ballmann2012}%
  \BibitemOpen
  \bibfield  {author} {\bibinfo {author} {\bibfnamefont {S.}~\bibnamefont
  {Ballmann}}, \bibinfo {author} {\bibfnamefont {R.}~\bibnamefont {H\"artle}},
  \bibinfo {author} {\bibfnamefont {P.~B.}\ \bibnamefont {Coto}}, \bibinfo
  {author} {\bibfnamefont {M.}~\bibnamefont {Elbing}}, \bibinfo {author}
  {\bibfnamefont {M.}~\bibnamefont {Mayor}}, \bibinfo {author} {\bibfnamefont
  {M.~R.}\ \bibnamefont {Bryce}}, \bibinfo {author} {\bibfnamefont
  {M.}~\bibnamefont {Thoss}},\ and\ \bibinfo {author} {\bibfnamefont {H.~B.}\
  \bibnamefont {Weber}},\ }\bibfield  {title} {\bibinfo {title} {Experimental
  evidence for quantum interference and vibrationally induced decoherence in
  single-molecule junctions},\ }\href
  {https://doi.org/10.1103/PhysRevLett.109.056801} {\bibfield  {journal}
  {\bibinfo  {journal} {Phys. Rev. Lett.}\ }\textbf {\bibinfo {volume} {109}},\
  \bibinfo {pages} {056801} (\bibinfo {year} {2012})}\BibitemShut {NoStop}%
\bibitem [{\citenamefont {Koch}\ \emph
  {et~al.}(2006{\natexlab{b}})\citenamefont {Koch}, \citenamefont {Semmelhack},
  \citenamefont {von Oppen},\ and\ \citenamefont {Nitzan}}]{Koch2006a}%
  \BibitemOpen
  \bibfield  {author} {\bibinfo {author} {\bibfnamefont {J.}~\bibnamefont
  {Koch}}, \bibinfo {author} {\bibfnamefont {M.}~\bibnamefont {Semmelhack}},
  \bibinfo {author} {\bibfnamefont {F.}~\bibnamefont {von Oppen}},\ and\
  \bibinfo {author} {\bibfnamefont {A.}~\bibnamefont {Nitzan}},\ }\bibfield
  {title} {\bibinfo {title} {Current-induced nonequilibrium vibrations in
  single-molecule devices},\ }\href
  {https://doi.org/10.1103/PhysRevB.73.155306} {\bibfield  {journal} {\bibinfo
  {journal} {Phys. Rev. B}\ }\textbf {\bibinfo {volume} {73}},\ \bibinfo
  {pages} {155306} (\bibinfo {year} {2006}{\natexlab{b}})}\BibitemShut
  {NoStop}%
\bibitem [{\citenamefont {H\"artle}\ and\ \citenamefont
  {Thoss}(2011{\natexlab{b}})}]{Haertle2011c}%
  \BibitemOpen
  \bibfield  {author} {\bibinfo {author} {\bibfnamefont {R.}~\bibnamefont
  {H\"artle}}\ and\ \bibinfo {author} {\bibfnamefont {M.}~\bibnamefont
  {Thoss}},\ }\bibfield  {title} {\bibinfo {title} {Vibrational instabilities
  in resonant electron transport through single-molecule junctions},\ }\href
  {https://doi.org/10.1103/PhysRevB.83.125419} {\bibfield  {journal} {\bibinfo
  {journal} {Phys. Rev. B}\ }\textbf {\bibinfo {volume} {83}},\ \bibinfo
  {pages} {125419} (\bibinfo {year} {2011}{\natexlab{b}})}\BibitemShut
  {NoStop}%
\bibitem [{\citenamefont {H\"artle}\ and\ \citenamefont
  {Kulkarni}(2015)}]{Haertle2015a}%
  \BibitemOpen
  \bibfield  {author} {\bibinfo {author} {\bibfnamefont {R.}~\bibnamefont
  {H\"artle}}\ and\ \bibinfo {author} {\bibfnamefont {M.}~\bibnamefont
  {Kulkarni}},\ }\bibfield  {title} {\bibinfo {title} {Effect of broadening in
  the weak-coupling limit of vibrationally coupled electron transport through
  molecular junctions and the analogy to quantum dot circuit qed systems},\
  }\href {https://doi.org/10.1103/PhysRevB.91.245429} {\bibfield  {journal}
  {\bibinfo  {journal} {Phys. Rev. B}\ }\textbf {\bibinfo {volume} {91}},\
  \bibinfo {pages} {245429} (\bibinfo {year} {2015})}\BibitemShut {NoStop}%
\bibitem [{\citenamefont {Gelbwaser-Klimovsky}\ \emph
  {et~al.}(2018)\citenamefont {Gelbwaser-Klimovsky}, \citenamefont
  {Aspuru-Guzik}, \citenamefont {Thoss},\ and\ \citenamefont
  {Peskin}}]{Gelbwaser2018}%
  \BibitemOpen
  \bibfield  {author} {\bibinfo {author} {\bibfnamefont {D.}~\bibnamefont
  {Gelbwaser-Klimovsky}}, \bibinfo {author} {\bibfnamefont {A.}~\bibnamefont
  {Aspuru-Guzik}}, \bibinfo {author} {\bibfnamefont {M.}~\bibnamefont
  {Thoss}},\ and\ \bibinfo {author} {\bibfnamefont {U.}~\bibnamefont
  {Peskin}},\ }\bibfield  {title} {\bibinfo {title} {High-voltage-assisted
  mechanical stabilization of single-molecule junctions},\ }\href
  {https://doi.org/10.1021/acs.nanolett.8b01127} {\bibfield  {journal}
  {\bibinfo  {journal} {Nano Lett.}\ }\textbf {\bibinfo {volume} {18}},\
  \bibinfo {pages} {4727} (\bibinfo {year} {2018})}\BibitemShut {NoStop}%
\bibitem [{\citenamefont {H\"artle}\ \emph {et~al.}(2018)\citenamefont
  {H\"artle}, \citenamefont {Schinabeck}, \citenamefont {Kulkarni},
  \citenamefont {Gelbwaser-Klimovsky}, \citenamefont {Thoss},\ and\
  \citenamefont {Peskin}}]{Haertle2018}%
  \BibitemOpen
  \bibfield  {author} {\bibinfo {author} {\bibfnamefont {R.}~\bibnamefont
  {H\"artle}}, \bibinfo {author} {\bibfnamefont {C.}~\bibnamefont
  {Schinabeck}}, \bibinfo {author} {\bibfnamefont {M.}~\bibnamefont
  {Kulkarni}}, \bibinfo {author} {\bibfnamefont {D.}~\bibnamefont
  {Gelbwaser-Klimovsky}}, \bibinfo {author} {\bibfnamefont {M.}~\bibnamefont
  {Thoss}},\ and\ \bibinfo {author} {\bibfnamefont {U.}~\bibnamefont
  {Peskin}},\ }\bibfield  {title} {\bibinfo {title} {Cooling by heating in
  nonequilibrium nanosystems},\ }\href
  {https://doi.org/10.1103/PhysRevB.98.081404} {\bibfield  {journal} {\bibinfo
  {journal} {Phys. Rev. B}\ }\textbf {\bibinfo {volume} {98}},\ \bibinfo
  {pages} {081404} (\bibinfo {year} {2018})}\BibitemShut {NoStop}%
\bibitem [{\citenamefont {Esposito}\ \emph {et~al.}(2009)\citenamefont
  {Esposito}, \citenamefont {Harbola},\ and\ \citenamefont
  {Mukamel}}]{Esposito2009}%
  \BibitemOpen
  \bibfield  {author} {\bibinfo {author} {\bibfnamefont {M.}~\bibnamefont
  {Esposito}}, \bibinfo {author} {\bibfnamefont {U.}~\bibnamefont {Harbola}},\
  and\ \bibinfo {author} {\bibfnamefont {S.}~\bibnamefont {Mukamel}},\
  }\bibfield  {title} {\bibinfo {title} {Nonequilibrium fluctuations,
  fluctuation theorems, and counting statistics in quantum systems},\ }\href
  {https://doi.org/10.1103/revmodphys.81.1665} {\bibfield  {journal} {\bibinfo
  {journal} {Rev. Mod. Phys.}\ }\textbf {\bibinfo {volume} {81}},\ \bibinfo
  {pages} {1665} (\bibinfo {year} {2009})}\BibitemShut {NoStop}%
\bibitem [{\citenamefont {Cerrillo}\ \emph {et~al.}(2016)\citenamefont
  {Cerrillo}, \citenamefont {Buser},\ and\ \citenamefont
  {Brandes}}]{Cerrillo2016}%
  \BibitemOpen
  \bibfield  {author} {\bibinfo {author} {\bibfnamefont {J.}~\bibnamefont
  {Cerrillo}}, \bibinfo {author} {\bibfnamefont {M.}~\bibnamefont {Buser}},\
  and\ \bibinfo {author} {\bibfnamefont {T.}~\bibnamefont {Brandes}},\
  }\bibfield  {title} {\bibinfo {title} {Nonequilibrium quantum transport
  coefficients and transient dynamics of full counting statistics in the
  strong-coupling and non-markovian regimes},\ }\href
  {https://doi.org/10.1103/PhysRevB.94.214308} {\bibfield  {journal} {\bibinfo
  {journal} {Phys. Rev. B}\ }\textbf {\bibinfo {volume} {94}},\ \bibinfo
  {pages} {214308} (\bibinfo {year} {2016})}\BibitemShut {NoStop}%
\bibitem [{\citenamefont {Jin}\ \emph {et~al.}(2015)\citenamefont {Jin},
  \citenamefont {Wang}, \citenamefont {Zheng},\ and\ \citenamefont
  {Yan}}]{Jin2015}%
  \BibitemOpen
  \bibfield  {author} {\bibinfo {author} {\bibfnamefont {J.}~\bibnamefont
  {Jin}}, \bibinfo {author} {\bibfnamefont {S.}~\bibnamefont {Wang}}, \bibinfo
  {author} {\bibfnamefont {X.}~\bibnamefont {Zheng}},\ and\ \bibinfo {author}
  {\bibfnamefont {Y.}~\bibnamefont {Yan}},\ }\bibfield  {title} {\bibinfo
  {title} {Current noise spectra and mechanisms with dissipaton equation of
  motion theory},\ }\href {https://doi.org/10.1063/1.4922712} {\bibfield
  {journal} {\bibinfo  {journal} {J. Chem. Phys.}\ }\textbf {\bibinfo {volume}
  {142}},\ \bibinfo {pages} {234108} (\bibinfo {year} {2015})}\BibitemShut
  {NoStop}%
\bibitem [{\citenamefont {Koch}\ \emph {et~al.}(2005)\citenamefont {Koch},
  \citenamefont {Raikh},\ and\ \citenamefont {von Oppen}}]{Koch2005FCS}%
  \BibitemOpen
  \bibfield  {author} {\bibinfo {author} {\bibfnamefont {J.}~\bibnamefont
  {Koch}}, \bibinfo {author} {\bibfnamefont {M.~E.}\ \bibnamefont {Raikh}},\
  and\ \bibinfo {author} {\bibfnamefont {F.}~\bibnamefont {von Oppen}},\
  }\bibfield  {title} {\bibinfo {title} {Full counting statistics of strongly
  non-ohmic transport through single molecules},\ }\href
  {https://doi.org/10.1103/PhysRevLett.95.056801} {\bibfield  {journal}
  {\bibinfo  {journal} {Phys. Rev. Lett.}\ }\textbf {\bibinfo {volume} {95}},\
  \bibinfo {pages} {056801} (\bibinfo {year} {2005})}\BibitemShut {NoStop}%
\bibitem [{\citenamefont {Golovach}\ and\ \citenamefont
  {Loss}(2004)}]{Golovach2004}%
  \BibitemOpen
  \bibfield  {author} {\bibinfo {author} {\bibfnamefont {V.~N.}\ \bibnamefont
  {Golovach}}\ and\ \bibinfo {author} {\bibfnamefont {D.}~\bibnamefont
  {Loss}},\ }\bibfield  {title} {\bibinfo {title} {Transport through a double
  quantum dot in the sequential tunneling and cotunneling regimes},\ }\href
  {https://doi.org/10.1103/PhysRevB.69.245327} {\bibfield  {journal} {\bibinfo
  {journal} {Phys. Rev. B}\ }\textbf {\bibinfo {volume} {69}},\ \bibinfo
  {pages} {245327} (\bibinfo {year} {2004})}\BibitemShut {NoStop}%
\bibitem [{\citenamefont {Gergs}\ \emph {et~al.}(2015)\citenamefont {Gergs},
  \citenamefont {H\"orig}, \citenamefont {Wegewijs},\ and\ \citenamefont
  {Schuricht}}]{Gergs2015}%
  \BibitemOpen
  \bibfield  {author} {\bibinfo {author} {\bibfnamefont {N.~M.}\ \bibnamefont
  {Gergs}}, \bibinfo {author} {\bibfnamefont {C.~B.~M.}\ \bibnamefont
  {H\"orig}}, \bibinfo {author} {\bibfnamefont {M.~R.}\ \bibnamefont
  {Wegewijs}},\ and\ \bibinfo {author} {\bibfnamefont {D.}~\bibnamefont
  {Schuricht}},\ }\bibfield  {title} {\bibinfo {title} {Charge fluctuations in
  nonlinear heat transport},\ }\href
  {https://doi.org/10.1103/PhysRevB.91.201107} {\bibfield  {journal} {\bibinfo
  {journal} {Phys. Rev. B}\ }\textbf {\bibinfo {volume} {91}},\ \bibinfo
  {pages} {201107} (\bibinfo {year} {2015})}\BibitemShut {NoStop}%
\bibitem [{\citenamefont {Kumar}\ \emph {et~al.}(2012)\citenamefont {Kumar},
  \citenamefont {Avriller}, \citenamefont {Yeyati},\ and\ \citenamefont {van
  Ruitenbeek}}]{Kumar2012}%
  \BibitemOpen
  \bibfield  {author} {\bibinfo {author} {\bibfnamefont {M.}~\bibnamefont
  {Kumar}}, \bibinfo {author} {\bibfnamefont {R.}~\bibnamefont {Avriller}},
  \bibinfo {author} {\bibfnamefont {A.~L.}\ \bibnamefont {Yeyati}},\ and\
  \bibinfo {author} {\bibfnamefont {J.~M.}\ \bibnamefont {van Ruitenbeek}},\
  }\bibfield  {title} {\bibinfo {title} {Detection of vibration-mode scattering
  in electronic shot noise},\ }\href
  {https://doi.org/10.1103/PhysRevLett.108.146602} {\bibfield  {journal}
  {\bibinfo  {journal} {Phys. Rev. Lett.}\ }\textbf {\bibinfo {volume} {108}},\
  \bibinfo {pages} {146602} (\bibinfo {year} {2012})}\BibitemShut {NoStop}%
\bibitem [{\citenamefont {Avriller}\ and\ \citenamefont
  {Frederiksen}(2012)}]{Avriller2012}%
  \BibitemOpen
  \bibfield  {author} {\bibinfo {author} {\bibfnamefont {R.}~\bibnamefont
  {Avriller}}\ and\ \bibinfo {author} {\bibfnamefont {T.}~\bibnamefont
  {Frederiksen}},\ }\bibfield  {title} {\bibinfo {title} {Inelastic shot noise
  characteristics of nanoscale junctions from first principles},\ }\href
  {https://doi.org/10.1103/PhysRevB.86.155411} {\bibfield  {journal} {\bibinfo
  {journal} {Phys. Rev. B}\ }\textbf {\bibinfo {volume} {86}},\ \bibinfo
  {pages} {155411} (\bibinfo {year} {2012})}\BibitemShut {NoStop}%
\bibitem [{\citenamefont {Viljas}\ \emph {et~al.}(2005)\citenamefont {Viljas},
  \citenamefont {Cuevas}, \citenamefont {Pauly},\ and\ \citenamefont
  {H\"afner}}]{Viljas2005}%
  \BibitemOpen
  \bibfield  {author} {\bibinfo {author} {\bibfnamefont {J.~K.}\ \bibnamefont
  {Viljas}}, \bibinfo {author} {\bibfnamefont {J.~C.}\ \bibnamefont {Cuevas}},
  \bibinfo {author} {\bibfnamefont {F.}~\bibnamefont {Pauly}},\ and\ \bibinfo
  {author} {\bibfnamefont {M.}~\bibnamefont {H\"afner}},\ }\bibfield  {title}
  {\bibinfo {title} {Electron-vibration interaction in transport through atomic
  gold wires},\ }\href {https://doi.org/10.1103/PhysRevB.72.245415} {\bibfield
  {journal} {\bibinfo  {journal} {Phys. Rev. B}\ }\textbf {\bibinfo {volume}
  {72}},\ \bibinfo {pages} {245415} (\bibinfo {year} {2005})}\BibitemShut
  {NoStop}%
\bibitem [{\citenamefont {Benito}\ \emph {et~al.}(2016)\citenamefont {Benito},
  \citenamefont {Niklas},\ and\ \citenamefont {Kohler}}]{Benito2016}%
  \BibitemOpen
  \bibfield  {author} {\bibinfo {author} {\bibfnamefont {M.}~\bibnamefont
  {Benito}}, \bibinfo {author} {\bibfnamefont {M.}~\bibnamefont {Niklas}},\
  and\ \bibinfo {author} {\bibfnamefont {S.}~\bibnamefont {Kohler}},\
  }\bibfield  {title} {\bibinfo {title} {Full-counting statistics of
  time-dependent conductors},\ }\href
  {https://doi.org/10.1103/PhysRevB.94.195433} {\bibfield  {journal} {\bibinfo
  {journal} {Phys. Rev. B}\ }\textbf {\bibinfo {volume} {94}},\ \bibinfo
  {pages} {195433} (\bibinfo {year} {2016})}\BibitemShut {NoStop}%
\end{thebibliography}%

\end{document}